\documentclass[twocolumn]{aastex7}

\usepackage{graphicx}	
\usepackage{amsmath}	
\usepackage{amssymb}	
\usepackage{afterpage}
\usepackage[T1]{fontenc}

\usepackage{CJKutf8}

\begin{document}

\begin{CJK}{UTF8}{gbsn}

\title{A close look at the black hole masses and hot dusty toruses of the first quasars with MIRI-MRS}

\author[orcid=0000-0001-8582-7012,sname='Bosman',gname='Sarah']{Sarah E.~I.~Bosman}
\affiliation{Institute for Theoretical Physics, Heidelberg University, Philosophenweg 12, 69120, Heidelberg, Germany}
\affiliation{Max-Planck-Institut f\"{u}r Astronomie, K\"{o}nigstuhl 17, 69117 Heidelberg, Germany}
\email[show]{bosman@thphys.uni-heidelberg.de}  

\author[orcid=0000-0002-7093-1877]{Javier \'{A}lvarez-M\'{a}rquez}
\affiliation{Centro de Astrobiolog\'{i}a (CAB), CSIC-INTA, Ctra. de Ajalvir km 4, Torrej \'{o}n de Ardoz, E-28850, Madrid, Spain}
\email{jalvarez@cab.inta-csic.es} 

\author[orcid=0000-0003-0821-3644]{Frederick B.~Davies}
\affiliation{Max-Planck-Institut f\"{u}r Astronomie, K\"{o}nigstuhl 17, 69117 Heidelberg, Germany}
\email{bosman@thphys.uni-heidelberg.de}  

\author[orcid=0009-0008-2205-7725]{Klaudia Protušová}
\affiliation{Institute for Theoretical Physics, Heidelberg University, Philosophenweg 12, 69120, Heidelberg, Germany}
\email{protusova@thphys.uni-heidelberg.de}  

\author[orcid=0000-0002-7054-4332]{Joseph F. Hennawi}
\affiliation{Leiden Observatory, Leiden University, P.O. Box 9513, 2300 RA Leiden, The Netherlands}
\affiliation{Department of Physics, University of California, Santa Barbara, CA 93106, USA}
\email{joe@physics.ucsb.edu}

\author[orcid=0000-0001-5287-4242]{Jinyi Yang}
\affiliation{Department of Astronomy, University of Michigan, 1085 S. University Ave., Ann Arbor, MI 48109, USA}
\email{yjykaren@gmail.com} 

\author[orcid=0000-0003-1634-1283]{Benedetta Spina}
\affiliation{Institute for Theoretical Physics, Heidelberg University, Philosophenweg 12, 69120, Heidelberg, Germany}
\email{spina@thphys.uni-heidelberg.de}


\author[orcid=0000-0002-9090-4227]{Luis Colina}
\affiliation{Centro de Astrobiolog\'{i}a (CAB), CSIC-INTA, Ctra. de Ajalvir km 4, Torrej \'{o}n de Ardoz, E-28850, Madrid, Spain}
\email{colina@cab.inta-csic.es}  

\author[orcid=0000-0003-3310-0131]{Xiaohui Fan}
\affiliation{Steward Observatory, University of Arizona, 933 N Cherry Avenue, Tucson, AZ 85721, USA}
\email{xiaohuidominicfan@gmail.com} 

\author[orcid=0000-0002-3005-1349]{G\"{o}ran \"{O}stlin}
\affiliation{Department of Astronomy, Stockholm University, Oscar Klein Centre, AlbaNova University Centre, 106 91 Stockholm, Sweden}
\email{ostlin@astro.su.se}  

\author[orcid=0000-0003-4793-7880]{Fabian Walter}
\affiliation{Max-Planck-Institut f\"{u}r Astronomie, K\"{o}nigstuhl 17, 69117 Heidelberg, Germany}
\email{walter@mpia.de}  

\author[orcid=0000-0002-7633-431X]{Feige Wang}
\affiliation{Department of Astronomy, University of Michigan, 1085 S. University Ave., Ann Arbor, MI 48109, USA}
\email{fgwang.astro@gmail.com}  

\author[orcid=0000-0003-1810-0889]{Martin Ward}
\affiliation{Centre for Extragalactic Astronomy, Durham University, South Road, Durham DH1 3LE, UK}
\email{martin.ward@durham.ac.uk}


\author[orcid=0000-0001-6794-2519]{Almudena Alonso Herrero}
\affiliation{Centro de Astrobiología (CAB), CSIC-INTA, Camino Bajo del Castillo s/n, E-28692 Villanueva de la Cañada, Madrid, Spain}
\email{aalonso@cab.inta-csic.es}

\author[0000-0002-3026-0562]{Aaron J.~Barth}
\affiliation{Department of Physics and Astronomy, 4129 Frederick Reines Hall, University of California, Irvine, CA, 92697-4575, USA}
\email{barth@uci.edu}

\author[orcid=0000-0003-4747-4484]{Silvia Belladitta}
\affiliation{Max-Planck-Institut f\"{u}r Astronomie, K\"{o}nigstuhl 17, 69117 Heidelberg, Germany}
\affiliation{INAF -- Osservatorio di Astrofisica e Scienza dello Spazio di Bologna, Via Gobetti 93/3, I-40129 Bologna, Italy}
\email{belladitta@mpia.de}

\author[orcid=0000-0002-3952-8588]{Leindert Boogaard}
\affiliation{Leiden Observatory, Leiden University, P.O. Box 9513, 2300 RA Leiden, The Netherlands}
\email{boogaard@strw.leidenuniv.nl}

\author[orcid=0000-0001-8183-1460]{Karina I.~Caputi}
\affiliation{Kapteyn Astronomical Institute, University of Groningen, P.O. Box 800, 9700 AV Groningen, The Netherlands}
\affiliation{Cosmic Dawn Center (DAWN), Denmark}
\email{karina@astro.rug.nl} 

\author[0000-0002-7898-7664]{Thomas Connor}
\affiliation{Center for Astrophysics $\vert$\ Harvard\ \&\ Smithsonian, 60 Garden St., Cambridge, MA 02138, USA}
\email{thomas.connor@cfa.harvard.edu}

\author[orcid=0000-0001-8986-5235]{Dominika Ďurovčíková}
\affiliation{Department of Physics, Massachusetts Institute of Technology, 77 Massachusetts Avenue, Cambridge, MA 02139, USA}
\affiliation{MIT Kavli Institute for Astrophysics and Space Research, 77 Massachusetts Avenue, Cambridge, MA 02139, USA}
\email{dominika.durovcikova@gmail.com}

\author[orcid=0000-0003-2895-6218]{Anna-Christina Eilers}
\affiliation{Department of Physics, Massachusetts Institute of Technology, 77 Massachusetts Avenue, Cambridge, MA 02139, USA}
\affiliation{MIT Kavli Institute for Astrophysics and Space Research, 77 Massachusetts Avenue, Cambridge, MA 02139, USA}
\email{eilers@mit.edu}

\author[orcid=0000-0003-2119-277X]{Alejandro Crespo Gómez}
\affiliation{Space Telescope Science Institute (STScI), 3700 San Martin Drive, Baltimore, MD 21218, USA}
\affiliation{Centro de Astrobiolog\'{i}a (CAB), CSIC-INTA, Ctra. de Ajalvir km 4, Torrej \'{o}n de Ardoz, E-28850, Madrid, Spain}
\email{acrespo@stsci.edu}

\author[orcid=0000-0002-4571-2306]{Jens Hjorth}
\affiliation{DARK, Niels Bohr Institute, University of Copenhagen, Jagtvej 155A, 2200 Copenhagen, Denmark}
\email{jens@nbi.ku.dk}

\author[orcid=0000-0003-1470-5901]{Hyunsung D.~Jun}
\affiliation{Department of Physics, Northwestern College, 101 7th St SW, Orange City, IA 51041, USA}
\email{hyunsung.jun@gmail.com} 

\author[orcid=0000-0001-5710-8395]{Danial Langeroodi}
\affiliation{DARK, Niels Bohr Institute, University of Copenhagen, Jagtvej 155A, 2200 Copenhagen, Denmark}
\email{danial.langeroodi@nbi.ku.dk}

\author[0000-0003-3762-7344]{Weizhe Liu (刘伟哲)}
\affiliation{Steward Observatory, University of Arizona, 933 N Cherry Avenue, Tucson, AZ 85721, USA}
\email{oscarlwz@gmail.com}

\author[0000-0001-6106-7821]{Alessandro Lupi}
\affiliation{Dipartimento di Scienza e Alta Tecnologia, Università degli Studi dell’Insubria, via Valleggio 11, 22100 Como, Italy}
\affiliation{INFN, Sezione di Milano-Bicocca, Piazza della Scienza 3, 20126 Milano, Italy}
\email{alessandro.lupi@unimib.it}

\author[orcid=0000-0002-5941-5214]{Chiara Mazzucchelli}
\affiliation{Instituto de Estudios Astrof\'{\i}sicos, Facultad de Ingenier\'{\i}a y Ciencias, Universidad Diego Portales, Avenida Ejercito Libertador 441, Santiago, Chile}
\email{chiara.mazzucchelli@mail.udp.cl}

\author[orcid=0000-0002-0932-4330]{John P.~Pye}
\affiliation{School of Physics \& Astronomy, Space Park Leicester, University of Leicester, 92 Corporation Road, Leicester LE4 5SP, UK}
\email{pye@leicester.ac.uk}

\author[orcid=0000-0002-5104-8245]{Pierluigi Rinaldi}
\affiliation{AURA for the European Space Agency (ESA), Space Telescope Science Institute, 3700 San Martin Dr., Baltimore, MD 21218, USA}
\affiliation{Steward Observatory, University of Arizona, 933 N Cherry Avenue, Tucson, AZ 85721, USA}
\email{prinaldi@stsci.edu}

\author[orcid=0000-0001-5434-5942]{Paul van der Werf}
\affiliation{Leiden Observatory, Leiden University, P.O. Box 9513, 2300 RA Leiden, The Netherlands}
\email{pvdwerf@strw.leidenuniv.nl}

\author[orcid=0000-0002-3216-1322]{Marta Volonteri}
\affiliation{Institut d'Astrophysique de Paris, Sorbonne Université, CNRS, UMR 7095, 98 bis bd Arago, 75014 Paris, France}
\email{martav@iap.fr}




\newpage

\begin{abstract}

The presence of supermassive black holes (SMBHs, $M_\text{BH} \sim 10^9 M_\odot$) at $z>7$ remains a puzzle.
While their existence appears to require exotic formation or growth processes, it is possible that BH mass estimates are incorrect due to differences with respect to the low-$z$ quasars on which BH mass scaling relations are calibrated.
In this work, we employ JWST MIRI-MRS spectroscopy to measure the rest-frame optical/IR properties of the four highest-redshift known luminous type-1 quasars at $7.08\leq z<7.64$. 
We use three new broad lines to measure updated BH masses, H$\alpha$, Pa$\alpha$ and Pa$\beta$, finding them to be in the range $(4-15) \cdot 10^8 M_\odot$. Our black hole mass estimates from all tracers agree with each other and with previous, less accurate, ground-based measurements based on Mg~{\small{II}}. The flux ratios of the H lines deviate from expectations for case A and B recombination in the same way as in $z<3$ quasars, indicating similar physical conditions in the Broad Line Region.
Rest-frame near-IR continuum emission from a hot dusty torus surrounding the accretion disc is unambiguously detected in all four objects. We model the emission with SKIRTOR and constrain the inclination (face-on) and the opening angle ($\theta=40-60^\circ$) of the tori. These constraints are consistent for the four objects and with expectations from luminous quasars. We estimate a total dust mass $(1-4) \cdot 10^6 M_\odot$ in the tori, corresponding to $(0.2-7)\%$ of the total dust in the quasar host galaxies. Given observed accretion rates, these SMBHs will deplete their tori in only $\sim5$ Myr.  
Overall, we confirm the fact that $z>7$ SMBHs in quasars could not have grown from stellar-remnant BHs if the radiative efficiency of accretion is $10\%$. We also find no evidence that inferred BH masses and accretion processes in $z>7$ quasars differ significantly from their near-identical counterparts at $z<3$.

\end{abstract}


\section{Introduction}

\begin{table*}
\label{tab1}
\hspace{-5em}\begin{tabular}{l c c c c c c}
\hline
\hline
Name & $z_\text{QSO}$ & $L_\text{bol}/L_\odot$ & MRS $t_\text{obs}/$s & MRS PID (PI) & NIRSpec PID (PI) & Discovery ref. \\
\hline
J0313$-$1806 & 7.642 & $3.6\cdot10^{13}$ & 3153 & 1764 (Fan) & 1764 (Fan) \& 3526 (Hennawi) & \citet{Wang21} \\
J1342+0928 & 7.541 & $4.0\cdot10^{13}$ & 3153 & 1219 (Lützgendorf) & 1219 (Lützgendorf) & \citet{Banados18} \\
J1007+2115 & 7.515 & $4.9\cdot10^{13}$ & 2484 & 1764 (Fan) & 1764 (Fan) \& 3526 (Hennawi) & \citet{Yang20} \\
J1120+0641 & 7.085 & $6.3\cdot10^{13}$ & 3153 & 1263 (Colina) & 1222 (Willott) & \citet{Mortlock11} \\
\hline
\hline
\end{tabular}
\caption{Data used in this work. MRS observations use three configurations (SHORT, MEDIUM, LONG) to cover the the full available wavelength range; the table lists the integration time per configuration. Redshifts are all obtained from the sub-mm C~[{\small{II}}] emission line of the host galaxies, and uncertainties on $L_\text{bol}$ are $\sim10\%$.} 
\end{table*}

The existence of supermassive black holes (SMBHs) with masses larger than $10^9 M_\odot$ less than a billion years after the Big Bang is one of the most compelling puzzles in astronomy. If our estimates of their masses are correct, it appears that the first SMBHs cannot have grown from the remnants of the first stars while maintaining rates of matter accretion below the classically-assumed Eddington limit (e.g.~\citealt{Inayoshi20}). This problem has motivated extensive theoretical work into ways producing massive ``seed'' black holes at $z\gtrsim10$ (e.g.~\citealt{Latif13, Luo18}) and into the possibility of BH growth channels where mass can be accreted at higher rates without giving rise to brightness in excess of the Eddington luminosity (``super-Eddington'' accretion; \citealt{Volonteri05,Lupi24}).

While the issues with growing overly-massive early SMBHs were first noticed in $z>6$ quasars \citep{Haiman01, Willott03,Volonteri06,Banados18}, comparably problematic masses are being inferred from the properties of so-called ``Little Red Dot'' AGN (LRDs) at even higher redshifts, $z\gtrsim10$ \citep{Matthee24,Maiolino24b,Taylor25,Napolitano25}. However, BH mass measurements in LRDs are controversial, as some of their observed properties differ greatly from the low-$z$ AGN on which BH mass estimators have been calibrated \citep{Rusakov25,Chang25}. In particular, LRDs appear to lack or be deficient in hot dust emission, a nearly universal component of AGN spectral energy distributions (SEDs) at later times \citep{Williams24,Setton25,Wang25}; LRDs also present extinction of their hydrogen broad-line region (BLR; e.g.~\citealt{Schindler24}), which is extremely rare in luminous quasars at $z<5$. While these unusual properties may point towards new SMBH accretion channels which might help explain large SMBH masses at $z>7$, they also call into question the suitability of current BH mass estimators for characterising LRDs.

In contrast, despite decades of scrutiny, very few changes (and no unique changes) have been observed in type-1 luminous quasars at $z>6$ compared to $0<z<4$. Their inferred BH masses of $10^{9-10}M_\odot$, although larger by $\sim1.5$ dex than those of quasars used to calibrate BH mass scaling relations, are not very different from the most massive BHs found at $z<3$ (e.~g.~compare \citealt{Wolf24} and \citealt{Wu15}). Their observed fluxes never exceed the Eddington luminosity by more than the scatter in single-epoch BH mass scaling relations \citep{Farina22,Mazzucchelli23}. Their radio-loudness \citep{Banados15,Keller24} and even the abundances of all accessible metals in their BLR \citep{Lai22,Jiang24} do not evolve with redshift compared with quasars of the same BH mass and accretion rates. Nevertheless, some signs of redshift evolution are present: quasars at $z>7$ are more likely to display extreme outflows ($v>5000$ km/s) in the ionised phase of their BLR \citep{Meyer19,Schindler20}, they are more likely to display broad absorption lines \citep{Bischetti22, Bischetti23} and they may be brighter in X-ray emission \citep{Zappacosta23,Tortosa24}. None of these changes are unique to early quasars, since they all individually fall within the scatter of $z<3$ quasars with comparable masses and luminosities. Nevertheless, a deeper investigation of potential redshift evolution is certainly warranted, both because it may be linked to changes in accretion properties at early times and because evolution may unknowingly bias determinations of BH masses.

In this context, we use mid-infrared spectroscopy of the $4$ highest-redshift currently-known quasars to explore both the robustness of single-epoch BH mass estimators and the presence and physical properties of the first BH accretion tori and BLR. We use the Medium Resolution Spectroscopy mode (MRS; \citealt{Wells+15,Argyriou+23}) of the Mid-Infrared Instrument (MIRI; \citealt{Rieke+15,Wright+15,Wright+23}) aboard the James Webb Space Telesope (JWST; \citealt{JWST,Gardner+23}) to cover the wavelength range $4.9<\lambda_\text{obs}<27.9$ $\mu$m with unprecedented sensitivity, as described in Section~\ref{sec:methods}.

First, in Section~\ref{sec:BL}, we use observations of the broad H$\alpha$, Pa$\alpha$ and Pa$\beta$ emission lines to update the BH mass measurements of the four highest-redshift quasars and investigate potential biases between tracers. Unlike the most commonly-used ground-accessible broad emission line used to estimate BH mass, Mg~{\small{II}} $\lambda 2798$\AA, the infrared H lines provide BH mass estimators based on the line luminosities directly rather than the continuum luminosity, and they more closely predict BH masses measured from reverberation mapping at late times. The Pa$\alpha$ and Pa$\beta$ transitions possess significantly longer wavelengths which should make them the least affected by potential dust extinction, the presence of which is also directly testable via emission line ratios. We summarise our results on the reliability of BH mass estimators in Section~\ref{sec:reliable} and discuss implications for the formation and growth of the first SMBHs in Section~\ref{sec:growth}.

Second, in Section~\ref{sec:torus}, we combine MRS spectra with complementary spectroscopy from NIRSpec  to conduct detailed modelling of the rest-frame near-IR SEDs of the four quasars, which all show unambiguously detected emission from hot dust in an accretion torus. Using a numerical model of the dusty torus, we constrain its physical parameters for the first time at $z>7$, including the structure's opening angle. We further discuss the physical consequences of our model, including estimating the torus depletion timescale, in Section~\ref{sec:dustmass}. In order to compare our results with quasars at lower redshifts, we discuss the use of a simple, widespread, one-temperature model of the hot torus in Section~\ref{sec:BB}; due to concerns about potential systematic biases, we also conduct a model-independent comparison of the $2\mu$m IR excess across redshift in Section~\ref{sec:excess}. We summarise our results in Section~\ref{sec:ccl}.

Throughout the paper we assume a $\Lambda$CDM cosmology with \textit{Planck} parameters $(h,\Omega_M) = (0.6774, 0.3089)$ \citep{Planck20} and distances are given in proper units.

\section{Methods}\label{sec:methods}

\begin{figure*}
\centering
\includegraphics[width=0.79\textwidth]{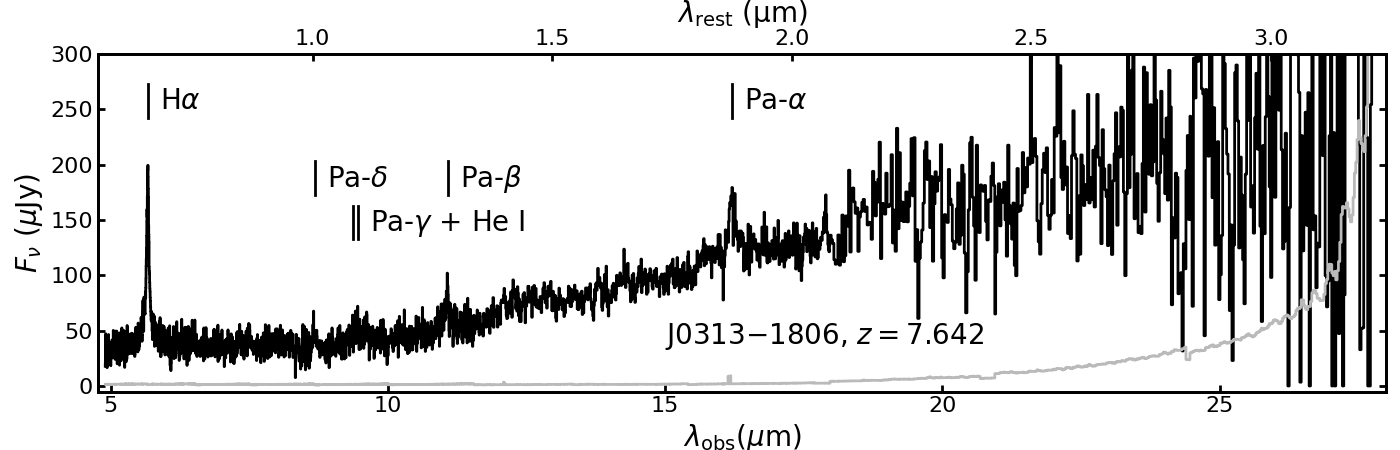}
\includegraphics[width=0.79\textwidth]{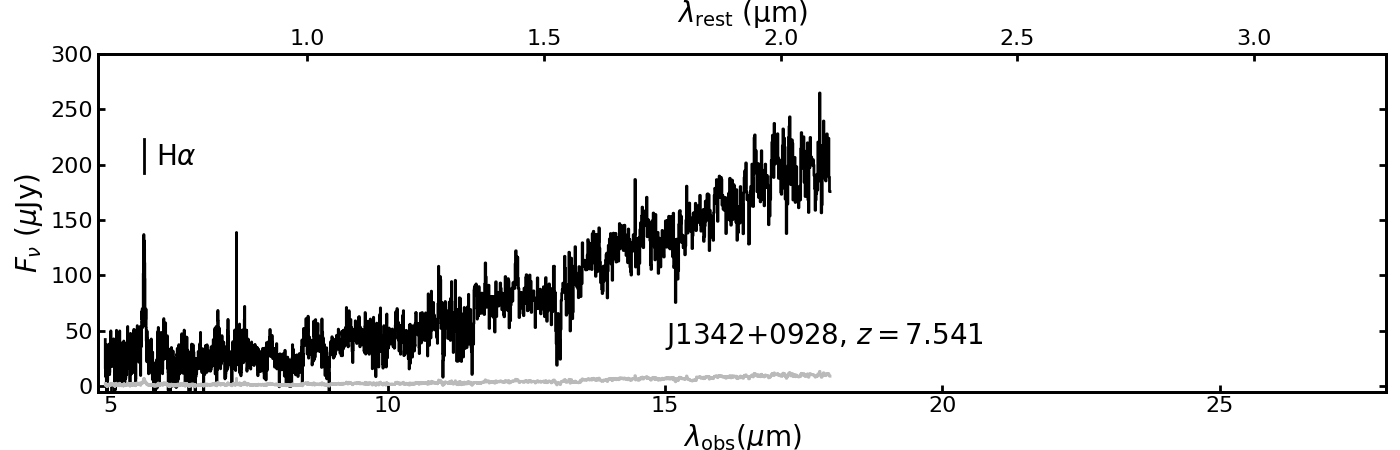}
\includegraphics[width=0.79\textwidth]{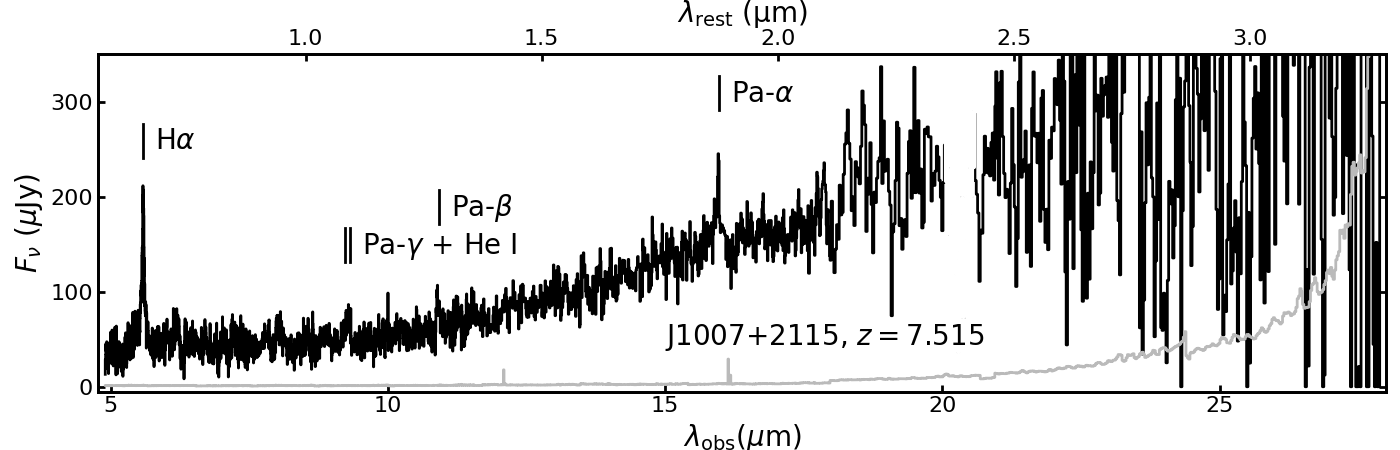}
\includegraphics[width=0.79\textwidth]{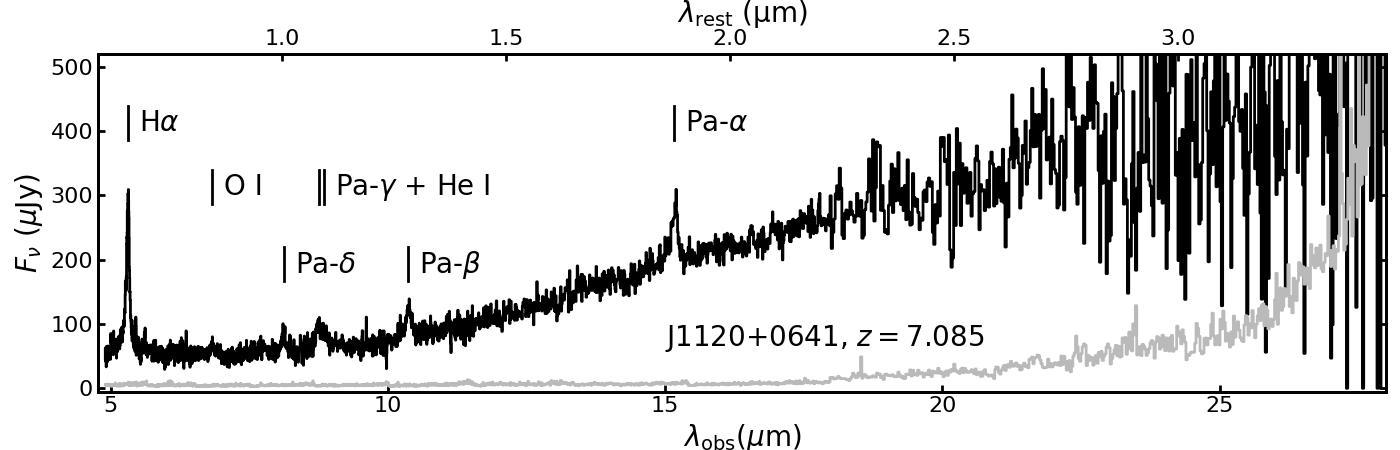}
\caption{MRS spectra of the four $z>7$ quasars in our sample; flux is in black and uncertainties are in grey. Channel 4 is not shown for J1342+0928 because the observations were badly affected by cosmic ray showers, prohibiting the extraction of a spectrum. Vertical lines denote the location of detected emission lines.}
\label{fig:all}
\end{figure*}

\subsection{MRS Spectroscopy}

Spectra were obtained with MRS the in 2023 as part of three observational programs: program IDs 1219 and 1263, as part of the guaranteed time awarded to the MIRI European Consortium, and the Cycle 1 GO program ID 1764, which aims to study $z>7.5$ across all wavelengths enabled by JWST (see e.g.~\citealt{Liu24,Pudoka25}). The integration times are similar for all targets (see Table~\ref{tab1}) and we reduced all spectra using the same procedure. Together, these four quasars are the four most-distant type-1 luminous quasars known in the literature at present.


MRS observations cover the wavelength range $4.9<\lambda<27.9$ $\mu$m, divided into $12$ sub-bands (CH1 SHORT, CH1 MEDIUM, CH1 LONG, $\dots$ , CH4 LONG). Each set of four SHORT sub-bands is observed simultaneously, as are the MEDIUM and LONG sub-bands. CH4 is significantly less sensitive than the other three channels by about a factor of $10$. The spectral resolution varies within and between sub-bands, with approximately $R\sim3500, 3000, 2500, 1500$ in CH1 through CH4 \citep{Labiano+21,Jones+23}. The spectra are uniformly sampled with $\Delta \lambda = 8, 13, 25, 60$\AA \ in CH1 through CH4, respectively. 

We generally follow the standard data-reduction pipeline for MRS observations \citep{Labiano+16,pipeline_mrs}. We use version 1.12.5 of the JWST calibration pipeline with version 11.17.7 and context 1149 of the Calibration Reference Data System (CRDS). The standard pipeline includes three broad steps: detector-level corrections and masking of bad pixels (including cosmic ray showers; \citealt{Argyriou+23}), calibrations using reference flat-fields, fringe flat and photometric reference files \citep{Patapis+23,Gasman+23}, and combination of all exposures within a sub-band into a 3D cube \citep{Law+23}. These steps are described in more detail in \citet{Bosman24}. While a newer version of the calibration pipeline has been released since the reductions were performed, we find no improvement in background subtraction or final SNR with the newer version, and therefore retain our analysis using the slightly older version.


Spectral extraction is performed in each of the MRS data cubes using circular apertures with radius $1\times$FWHM($\lambda$), where FWHM$(\lambda) = 0.3''$ at $\lambda < 8\mu$m and FWHM$(\lambda) = 0.31'' \times \lambda(\mu\text{m})/8\mu\text{m}$ at $λ\lambda 8\mu$m, following the FWHM of the MRS Point Spread Function (PSF). Aperture losses are accounted for within the pipeline.

Finally, the set of 12 sub-bands is stitched into a single spectrum for each quasar. There is no evidence of mismatch of the flux calibration between the sub-bands in the overlapping wavelength ranges, and therefore no further normalisation is required. In the overlapping wavelength ranges, the fluxes are averaged and the uncertainties combined in quadrature. The final reduced and stitched spectra are shown in Figure~\ref{fig:all}.

During the observation of one of the targets, J1342+0928, the MRS experienced an above-average occurrence rate of cosmic ray (CR) events. These CRs produced extensive and persistent shower features in the detector, to the extent that over half of all spaxels in CH4 were visibly affected (worsened by the lower spatial sampling). As a result, the quasar is not convincingly detected in the three longest wavelength sub-bands which make up CH4, and we were unable to extract a spectrum. As can be seen in Figure~\ref{fig:all}, the CR residuals produce significant artifacts also in CH1-CH3, and only the H$\alpha$ emission line could be detected with certainty. All raw and calibration data used in this work from both MIRI and NIRSpec may be obtained from the online Mikulshi Archive for Space Telescopes\footnote{\url{https://archive.stsci.edu/}} (MAST) by using the PIDs listed in Table~\ref{tab1}.

\subsection{NIRSpec spectroscopy}\label{sec:nirspec}

For the purposes of modelling the rest-frame infrared emission from the dusty torus self-consistently, as will be described in Section~\ref{sec:torus}, we require a measurement of the UV and optical power-law slopes of the accretion disc continuum. The optical accretion disc continuum slope is needed for statistically meaningful fitting of a joint accretion disc and torus model over the wavelength range covered by the MRS, while the UV slope is needed to provide an estimate of the total luminosity of the accretion disc, which gets re-processed by dust in the torus. To measure these slopes, we employ NIRSpec spectroscopy \citep{NIRSpec,NIRSpec2} of the quasars in the rest-frame UV and optical obtained through various observational programs, as listed in Table \ref{tab1}. Further analysis of the NIRSpec spectra to measure the rest-frame optical properties of the quasars will be presented in Yang et al.~(in prep.).

J0313$-$1806 and J1007+2115 were observed in fixed-slit configuration in the high-resolution G140H, G235H and G395H gratingss for 5252 seconds of integration time each as part of the same program as their MRS spectroscopy (PID 1764). Both quasars were later re-observed for 12744 seconds ($\sim3.54$ hours) in the G235H grating only as part of program ID 3526, which aims to detect the presence of the so-called ``Mg~{\small{II}} forest'' in their foreground \citep{Hennawi21}. 
J1342+0928 and J1120+0641 were observed with the NIRSpec Micro-Shutter Assembly (MSA) multi-slit spectrograph in the G140H and G235H gratings as part of GTO programs; the quasars were placed in the Fixed Slits of the MSA (see also \citealt{Christensen23,Kist25}). All NIRSpec spectra were reduced using a combination of the PypeIt reduction pipeline \citep{Pypeit-official} and the officia  JWST Science calibration pipeline CALWEBB (version 1.13.4). Further details will be provided in Hennawi et al.~(in prep).

\begin{table*}[htb!]
\centering
\begin{tabular}{l c c c c}
\hline
\hline
& J0313$-$1806 & J1342+0928 & J1007+2115 & J1120+0641 \\
\hline
FWHM${}_{\text{H}\alpha}$ (km s${}^{-1}$) & $2690 \pm 110$ & $2340\pm 90$ & $3075 \pm 160$ & $3410 \pm 120$ \\
(BC1, km s${}^{-1}$) & $2230\pm 30$  & $-$ & $1860\pm 70$ & $2920 \pm 80$ \\
(BC2, km s${}^{-1}$) & $12100\pm 260$ & $-$ & $6830\pm 470$ & $12200\pm 900$ \\
$F_{\text{H}\alpha}$ ($10^{-16}$ cgs) & $13.98 \pm 0.11$ & $6.6 \pm 1.0 $ & $10.2^{+1.9}_{-2.0}$ & $24.5\pm 2.0$ \\
\hline
FWHM${}_{\text{Pa}\alpha}$ (km s${}^{-1}$) & $2490\pm 90$ & $-$ & $2259\pm 160$ & $1980 \pm 210$ \\
(BC1, km s${}^{-1}$) & $-$&$-$ &$-$ & $1570 \pm 180$ \\
(BC2, km s${}^{-1}$) &$-$ &$-$ &$-$ & $3270 \pm 390$ \\
$F_{\text{Pa}\alpha}$ ($10^{-16}$ cgs) & $0.82 \pm 0.02$ & $-$ & $0.80 \pm 0.22$ & $2.30\pm 0.31$ \\
\hline
FWHM${}_{\text{Pa}\beta}$ (km s${}^{-1}$) & $3320\pm 180$ & $-$ & $2570\pm 60$ & $3430\pm 230$ \\
$F_{\text{Pa}\beta}$ ($10^{-16}$ cgs) & $1.23 \pm 0.19$ & $-$ & $ 0.49 \pm 0.10$ & $1.62\pm 0.15$ \\
\hline
\hline
\end{tabular}
\caption{Emission lines properties of the four quasars, listing the full width at half maximum (FWHM) and line fluxes. Fluxes are given in units of erg s$^{-1}$ cm$^{-2}$ (cgs). For some lines where two Gaussian components were used, the FWHM of both components are listed (BC1 and BC2).}
\label{tab2}
\end{table*}

\subsection{Emission line measurements}

Multiple broad emission lines are detected in the MRS spectra, including H$\alpha$ 6564\AA, Pa$\alpha$ 1.8756 $\mu$m, Pa$\beta$ 1.2822 $\mu$m, and the blend of Pa$\gamma$ 1.0941 $\mu$m with He~{\small{I}} 1.0833 $\mu$m. The first three of these emission lines have been calibrated, with various degrees of directness, to serve as proxies of the BH mass measured through reverberation-mapping (RM). To check agreement between different BH mass tracers, we therefore wish to measure the emission line fluxes, full widths at half maxima (FWHM), and centres.




\begin{figure*}[th]
\centering
\includegraphics[width=\textwidth]{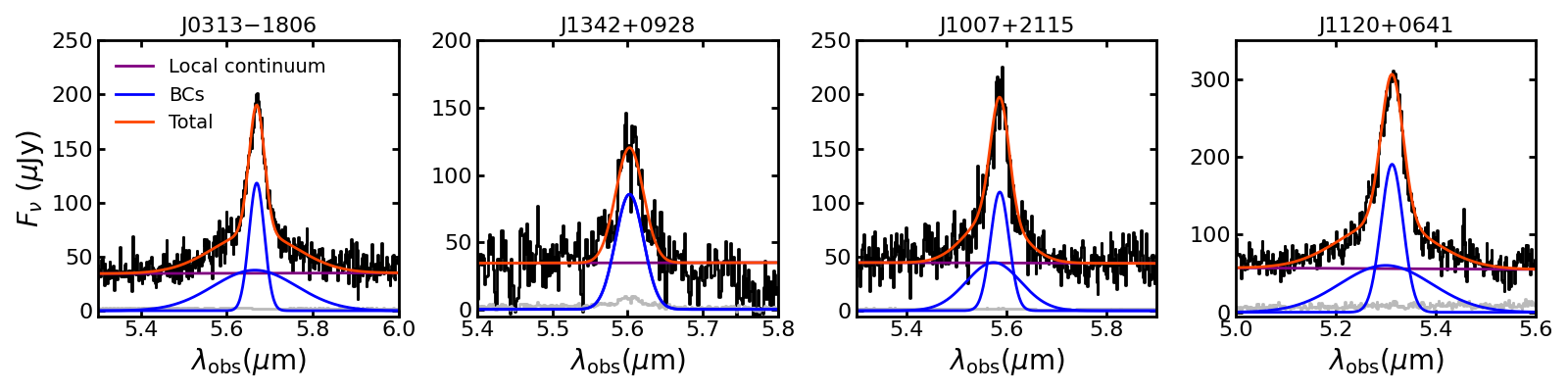}
\caption{Broad H$\alpha$ lines detected in the four $z>7$ quasars. Blue shows the broad components (BCs). The broadest line components for J0313$-$1806 and J1120+0641 have FWHM$\sim12000$ km s${}^{-1}$. In J1342+0928, CR shower residuals affected the observations (see e.~g.~around $\lambda_\text{obs} = 5.8\mu$m) and only one broad component could be identified.}
\label{fig:ha}
\end{figure*}

We perform emission-line fitting using the publicly-available code {\texttt{SCULPTOR}} \citep{sculptor}. The parameters describing the emission lines and the local continuum emission are fit simultaneously. We model the local continuum as a power-law and the lines with either 1 or 2 broad Gaussian components. The spectral SNR is not sufficient to clearly detect emission from the Fe~{\small{II}} continuum, which is also much weaker in the rest-frame IR than the rest-frame optical. The fitting window includes 20 pixels of continuum emission on either side of the emission line. All parameters of the line and continuum fit are varied simultaneously; we sample the resulting 5 or 8 parameter space with 10,000 draws of a Monte Carlo Markov Chain (MCMC) sampler. For the sole emission line detected in quasar J1342+0928, we adjust the fitting window manually to avoid residuals from CR showers. This was not necessary for other emission lines. The measurements of FWHM and line centres, and their uncertainties, are obtained by sampling from the posterior distribution of the Gaussian component(s) of each line. No parameters are tied between lines. Resulting measurements are given in Table~\ref{tab2}. For the lines which required a double-Gaussian fit, the FWHM of the two components are also listed. The resulting fits for the H$\alpha$ emission lines are shown in Figure~\ref{fig:ha}.

\begin{figure*}[th]
\centering
\includegraphics[width=0.85\textwidth]{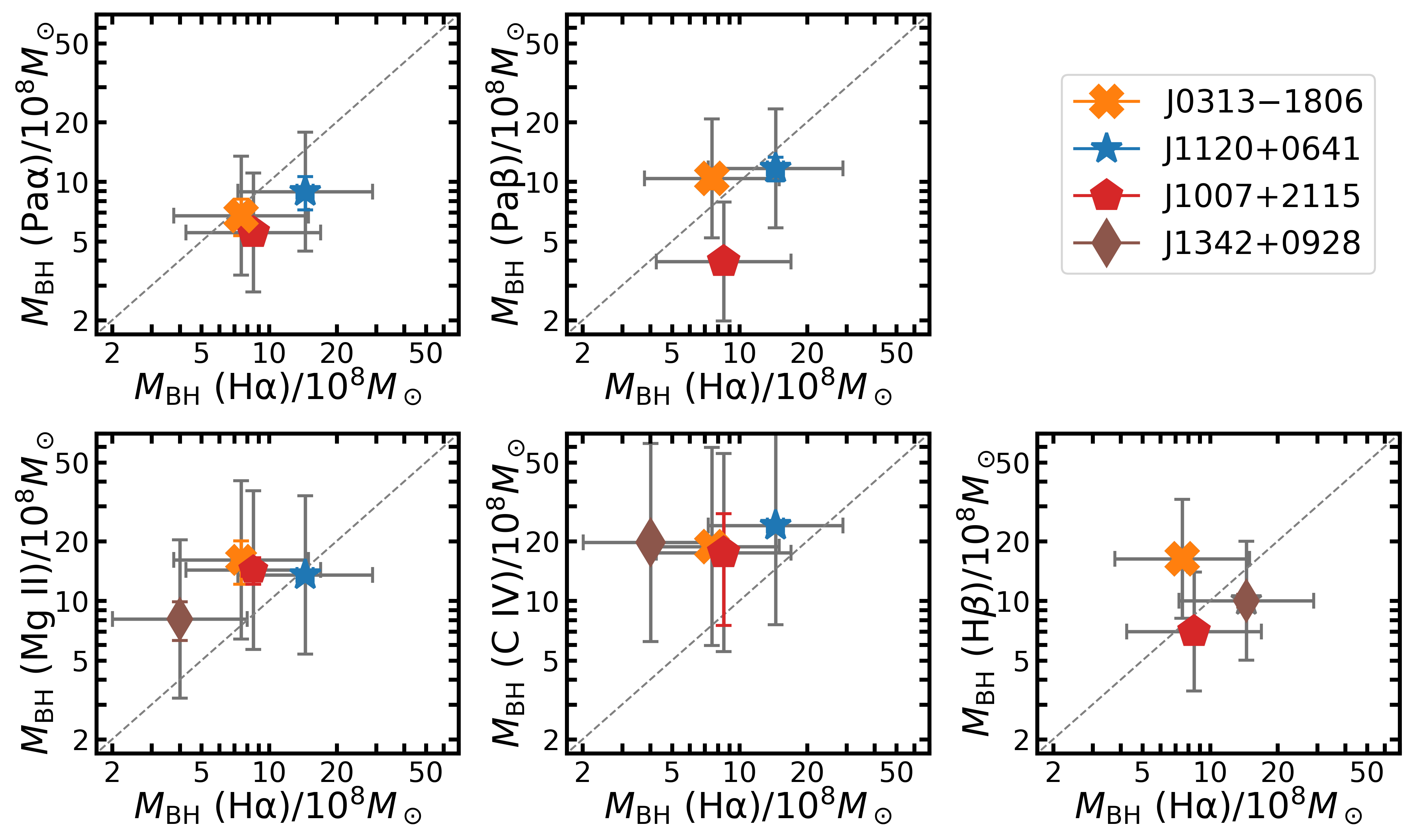}
\caption{Black hole masses calculated from five different single-epoch virial mass estimators: H$\alpha$, Pa$\alpha$ and Pa$\beta$ (this work), C~{\small{IV}} and Mg~{\small{II}} (this work, \citealt{Yang21,Farina22}). Measurement uncertainties are shown as coloured error bars while grey error bars reflect the intrinsic (systematic) scatter in the mass estimators. Despite implementing a correction of blueshift \citep{Coatman17}, C~{\small{IV}} remains a poor and potentially biased estimator. The other tracers agree within scatter, indicating no bias from ground-based measurements of the Mg~{\small{II}} emission line.}
\label{fig:bhm}
\end{figure*}

The measurement method for broad line properties for quasar J1120+0641 differs from the approach used in \citet{Bosman24}, where functional forms for all emission lines were fit together with the parameters describing continuum emission. Since our model for continuum emission is significantly more complex than the single-component black-body emission employed in the previous paper (see Section~\ref{sec:torus}), this was not practical for our study. As a result, we re-measure the emission line properties of J1120+0641 with a local continuum. We note that none of the new measurements differ from the previous ones by more than 1.5$\sigma$.

\section{Broad emission lines}\label{sec:BL}

\subsection{BH masses}

We calculate the BH masses of the quasars using single-epoch virial mass estimators calibrated on the FWHM and luminosities of the H$\alpha$, Pa$\alpha$, and Pa$\beta$ emission lines. We compare the results with the BH masses obtained using the C~{\small{IV}} $\lambda 1549$\AA \ and Mg~{\small{II}} emission lines from ground-based observations of the rest-frame UV in order to assess whether pre-JWST measurements may have been biased by extinction.

Virial mass estimators based on the properties of the H$\alpha$ line have been determined by various authors, relying on the observation that the widths of the H$\alpha$ and H$\beta$ emission lines trace each other closely. In this work, we adopt the best-fit scaling relation from \citet{Shen11} based on \citet{Greene05}:
\begin{equation}\label{eq:ha}
\begin{aligned}
\log M_{\text{BH, H}\alpha} = 0.379 + 0.43 \log \left(\frac{L_{\text{H}\alpha}}{10^{42} \text{erg s}^{-1}} \right) \\
+ 2.1 \log \left( \frac{\text{FWHM}_{\text{H}\alpha}}{\text{km s}^{-1}}\right).
\end{aligned}
\end{equation}

The Paschen-series emission lines have been calibrated as virial mass estimators by \citet{Kim10} and \citet{Kim15-Pa}, who used as reference values a mixture of BH mass measurements from reverberation mapping but also indirect measurements via the single-epoch estimator for the H$\beta$ line. These scaling relations have a reported intrinsic scatter compared to tracers from other emission lines of $\lesssim 0.2$ dex, but we inflate the estimated systematic uncertainty to $0.35$ dex to reflect the fact that the best H$\beta$ single-epoch estimators have, by themselves, a scatter of $\sim0.3$ dex compared to measurements from reverberation mapping \citep{Shen11}. The two single-epoch mass recipes we use are:
\begin{equation}\label{eq:pa}
\begin{aligned}
M_{\text{BH, Pa}\alpha} = 10^{7.31} \left(\frac{L_{\text{Pa}\alpha}}{10^{42} \text{erg s}^{-1}} \right)^{0.48} \\
\left( \frac{\text{FWHM}_{\text{Pa}\alpha}}{10^3 \text{ km s}^{-1}}\right)^{1.68},
\end{aligned}
\end{equation}
and
\begin{equation}\label{eq:pb}
\begin{aligned}
M_{\text{BH, Pa}\beta} = 10^{7.04} \left(\frac{L_{\text{Pa}\beta}}{10^{42} \text{erg s}^{-1}} \right)^{0.48} \\
\left( \frac{\text{FWHM}_{\text{Pa}\beta}}{10^3 \text{ km s}^{-1}}\right)^{2}.
\end{aligned}
\end{equation}

We further gather the ground-based BH mass measurements based on the Mg~{\small{II}} emission line from \citet{Yang21}. Their measurements employ the estimator from \citet{Vestergaard09}, which displays intrinsic scatter compared to BH masses measured via reverberation mapping of $\sim0.4$ dex. Unlike the single-epoch BH mass estimators based on H emission lines, the one based on Mg~{\small{II}} employs the luminosity of the accretion disc directly rather than relying on the luminosity of the emission line itself. Similarly, we gather BH mass measurements based on the H$\beta$ line obtained from the NIRSpec data of three of the quasars (\citealt{Trefoloni25,Liu24};Wolf et al.~submitted) and refer interested readers to those publications for details of the relations used. We note that \citet{Trefoloni25} do not provide measurement-only uncertainties for J1342+0928, and therefore we use only the systematic uncertainties of $\sim 0.3$ dex in Fig.~\ref{fig:bhm}. The BH mass from H$\beta$ for the last quasar, J1120+0641, will be presented in Yang et al.~(in prep). 

\begin{table*}[htb!]
\label{tab:bhm}
\begin{centering}
\hspace{-5em}\begin{tabular}{l c c c c c c l}
\hline
\hline
 & $M_{\text{BH, H}\alpha}$ & $M_{\text{BH, Pa}\alpha}$ & $M_{\text{BH, Pa}\beta}$ & $M_{\text{BH, Mg~II}}$ & $M_{\text{BH, C~IV}}$ & $M_{\text{BH, H}\beta}$ & Refs. \\
\hline
J0313$-$1806 & $7.52\pm 0.66$ & $6.75\pm 1.41$ & $10.4\pm 1.2$ & $16.1\pm 4.0 $ & $18.8\pm 1.8$ & $16.3\pm 1.0$ & 1,1,1,2,1,4  \\
J1342+0928 & $4.01 \pm 0.27$ & $-$ & $-$ & $8.1 \pm 1.8$ & $19.7 \pm 0.8$ & $10.0 $ & 1,$-$,$-$,2,3,5 \\
J1007+2115 & $8.5\pm 0.9$ & $ 5.55 \pm 0.67 $ & $3.95 \pm 0.24$ & $ 14.3 \pm 2.2$ & $ 17.5 \pm 10.0$ & $7.0 \pm 0.4$ & 1,1,1,2,1,6 \\
J1120+0641 & $14.5 \pm 1.2$ & $8.9\pm 1.7$ & $ 11.7\pm 1.6$ & $13.5\pm 0.4$ & $24.0 \pm 0.6$ & $-$ & 1,1,1,2,3,$-$ \\
\hline
\hline
\end{tabular}
\end{centering}
\caption{SMBH masses measured with single-epoch estimators based on five different emission lines. All masses are in units of $10^8 M_\odot$. Measurement uncertainties are given, but systematic uncertainties are omitted; the scaling relations have precisions of $0.3,0.35,0.35
,0.4,0.5$ dex for the H$\alpha$, Pa$\alpha$, Pa$\beta$, Mg~{\small{II}} and C~{\small{IV}} lines, respectively. References: [1] This work; [2] \citet{Yang21}; [3] \citet{Farina22}; [4] Wolf et al.~submitted; [5] \citet{Trefoloni25}; [6] \citet{Liu24}. }
\end{table*}

Finally, we calculate the BH masses estimated from the C~{\small{IV}} line based on measurements of the emission lines reported in the literature \citep{Farina22}. Since the C~{\small{IV}} emission line is accessible in the visible wavelengths up to $z\sim5.5$ unlike all other lines discussed thus far, it has been a convenient choice for estimating BH masses in large samples of quasars (e.g.~\citealt{Wu22}). Unfortunately, C~{\small{IV}} is known to display increasingly common extreme blueshifts at $z>6$ \citep{Meyer19}, and these blueshifts are known to quantitatively affect the virial mass estimator. To account for this bias, we use the virial mass formula of \citet{Coatman17}, which is designed to account for blueshift as an effective correction to the FWHM:
\begin{equation}\label{eq:civ}
\begin{aligned}
M_{\text{BH, CIV}} = 10^{6.71} \left(\frac{\lambda L_\lambda}{10^{44} \text{erg s}^{-1}} \right)^{0.53} \\
\left( \frac{\text{FWHM}_{\text{CIV, corr}}}{10^3 \text{ km s}^{-1}}\right)^{2},
\end{aligned}
\end{equation}
where
\begin{equation}
\begin{aligned}
\text{FWHM}_{\text{CIV, corr}} = \text{FWHM}_{\text{CIV}}  \times \\
\left( 0.41 \left(\frac{\Delta v_\text{CIV}}{10^3 \text{ km s}^{-1}} \right) +0.62 \right)^{-1}.
\end{aligned}
\end{equation}

The virial mass estimates based on Pa$\alpha$ and Pa$\beta$ make the underlying assumption that $R_\text{BLR} \propto L_{5100}^b$ with $b=0.52$, while all other estimates used here assume $b=0.50$. This leads to the estimates based on Paschen lines to be systematically over-estimated by $\sim 3-5\%$, which is sub-dominant to other sources of uncertainties and we therefore  neglect in the rest of the discussion.

We report all estimated BH masses in Table~\ref{tab:bhm} and show their agreement in Figure~\ref{fig:bhm}. We find that only the C~{\small{IV}}-based measurements of BH mass deviate systematically from tracers based on other lines; the BH mass measurements based on C~{\small{IV}} are highly similar for all four quasars and always larger than estimates based on H$\alpha$ by factors of $2-5$. Given the known lower accuracy of the C~{\small{IV}} mass estimator, this may reflect a residual bias of the tracer for line blueshifts in excess of the ones accounted for in \citet{Coatman17}. 

The tracers based on H$\alpha$, Pa$\alpha$ and Pa$\beta$ have lower systematic scatter than the ones based on the Mg~{\small{II}} emission line, and all converge on the assessment that J1120+0641 hosts the most massive known SMBH at $z>7$ ($M_\text{BH} \gtrsim 10^9 M_\odot$) while the other three quasars have comparable but slightly lower BH masses, $M_\text{BH} \sim 5 \cdot 10^8 M_\odot$.

\subsection{Broad line region}

The flux ratios of the broad hydrogen emission lines in quasars are known not to follow expectations from classical case A $-$ recombination of hydrogen through all hydrogen states $-$ nor case B  $-$ recombination through all states except the ground state \citep{Osterbrock89,Osterbrock06}. However, these ratios are well-explained by relatively simple equilibrium models of a slab of gas illuminated by incident radiation from an accretion disk \citep{Oyabu09}. To examine the broad-line flux ratios of our quasars, we re-use the \texttt{Cloudy} \citep{Cloudy} models from \citet{Bosman24}, which were themselves a simplified version of the model of \citet{Tsuzuki06} as implemented by \citet{Oyabu09} $-$ compared to those authors, we neglected metalicity and turbulence, but recover the same trends in emission line ratios. 

In this framework, the BLR is modelled as a single cloud of gas with density $n_\text{H}$ illuminated by a central source (see e.g.~\citealt{Goad12}). The incident radiation is given by a SED in the form of a UV bump combined with a power-law of X-ray emission and a cut-off temperature $T=150,000$ K. The intensity of the incident radiation is described by the ionisation parameter $U = \Phi/cn_\text{H}$, where $\Phi \ (\text{s}^{-1} \text{cm}^{-2})$ is the photon flux. We sample a parameter grid $\log n_\text{H} \ (\text{cm}^{-3}) = [10, 12, 14]$ and $-7<\log U<0$ in steps of $\Delta \log U = 1$. We refer interested readers to the papers above for more details. We note that our model is mostly used for qualitative purposes, in order to illustrate that the observed hydrogen  emission line ratios can arise from physical conditions which can plausibly arise in a quasar broad-line region (BLR), without the presence of extinction by dust or gas. In reality, the effects of turbulence, non-equilibrium hydrodynamics, magnetic fields, and metalicity cannot be quantitatively neglected (e.g.~\citealt{Krause12,Lattimer24}) and more careful modelling would be required for precise inference of BLR physical conditions.

\begin{figure}[t]
\includegraphics[width=\columnwidth]{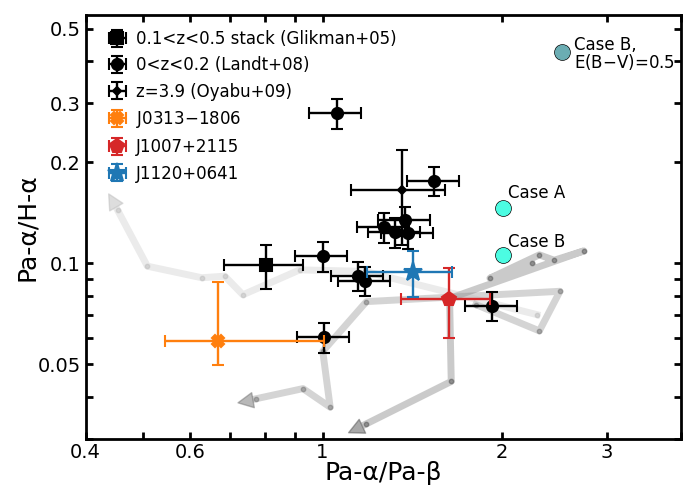}
\caption{Flux ratios of the broad H$\alpha$, Pa$\alpha$ and Pa$\beta$ lines. Tracks show the output of a simple \texttt{Cloudy} model with parameters of broad line region density $\log n_\text{H} (\text{cm}^{-3}) = 10, 12, 14$ from light to dark, and ionisation parameter increasing in the direction of the arrows over $-7<\log U<0$. Hydrogen BLRs are not consistent with case B nor case A recombination, but not in a direction which would correspond to extinction: the dark blue point shows expectations for case B with E$($B$-$V$)=0.5$.}
\label{fig:ratio}
\end{figure}

Nevertheless, we find that the ratios of H$\alpha$, Pa$\alpha$ and Pa$\beta$ fluxes are well captured by this simple model. While all three quasars for which measurements are possible deviate significantly from expectations from case A and case B recombination, they deviate in the same general direction as other quasars and AGN at lower redshifts \citep{Glikman06,Landt08,Oyabu09} and are consistent with our grid of models with an extinction E(B$-$V)$=0$ (Figure \ref{fig:ratio}). Quasars J1120+0641 and J0313$-$1806 lie closer to the models corresponding to $\log n_\text{H} (\text{cm}^{-3}) = 14$, while quasar J1007+2115 is well-explained by all models.

J0313$-$1806 displays an unusually weak Pa$\alpha$ emission, less luminous than the Pa$\beta$ line at $1\sigma$; in our model, this can be explained by an ionisation parameter stronger than for the other two quasars by a factor of $\sim 100$. We note that, within $1\sigma$ uncertainties, this quasar is not an outlier compared to lower-redshift objects. J0313$-$1806 displays by far the fastest BAL outflows out of any quasar at $z>7$, reaching at least 18\% of the speed of light \citep{Wang21}. Quasars with such fast outflows are extremely rare: for example, \citet{Rodriguez-Hidalgo20} found only $40$ quasars with BAL outflows faster than $0.1c$ out of $\sim6700$ luminous quasars at $2<z<5$; such outflows become more common at $z>6$ \citep{Bischetti22, Bischetti23,Belladitta25}. The extreme outflow speed may therefore be connected to unusually strong photon flux onto the BLR, reflected by a high $\log U$ in our model. However, we warn against an overly quantitative interpretation of this result, due to important modelling caveats mentioned above.

\section{Hot dusty torus}\label{sec:torus}

A prominent feature of AGN is an upturn in emission at $\lambda_\text{rest} \gtrsim 1\mu$m, usually attributed to emission from hot dust in a surrounding torus due to its close resemblance to black-body emission with $T\sim1200$ K \citep{Glikman06, Jun13}. The upturn is clearly detected in all four quasars in our sample as a change in the slope of the continuum at $\lambda_\text{obs}>10\mu$m compared to $\lambda_\text{obs}<10\mu$m (Fig.~\ref{fig:all}). 

Previous observational studies of hot dust in luminous quasars have often modeled the emission as a (sometimes modified) black-body of a single temperature \citep{Leipski14,Hernan16,Bosman24}. The underlying assumption is that the hottest component of the dusty torus dominates the observed emission due to the much steeper scaling of black-body emission with temperature compared to dust mass. However, this simplifying assumption is known to be insufficient for describing the mid-IR emission of fainter AGN. Instead, the dust density and temperature distributions inside the torus are modeled as continuous (with radial and polar gradients in density and temperature) and/or clumpy (to varying degrees); the latter assumption may better explain the large variation among AGN in the strength in dust silicate features observed at $\lambda_\text{rest}=9.7 \mu$m (e.g.~\citealt{Krolik88,Nenkova02,Dullemond05,Nikutta09,Honig10,Feltre12,Xu20,Garcia-Bernete24}). 

A few studies have similarly employed models including both clumpy and diffuse dust components \citep{Mor09,Mor12} to model IR emission in type-1 AGN, but the objects they considered had bolometric luminosities on average $100$ times fainter than ours; even their brightest targets were $\sim 5$ times fainter than our faintest quasar. To move beyond a single-component description of the torus in the most luminous quasars and constrain additional physical properties of the torus, we use the SKIRTOR model, a numerical simulation suite of the AGN dusty torus as a clumpy two-phase medium.

\subsection{Modelling with SKIRTOR}\label{model}

SKIRTOR employs radiative transfer to consistently model the re-processing of emission from a central AGN accretion disc through a surrounding torus of dust \citep{Stalevski12,Stalevski16}. The model employs the radiative transfer code SKIRT with the assumption of local thermal equilibrium \citep{Baes03, Baes11}. The dust consists of a mixture of graphite and silicate grains with the former being slightly less abundant \citep{Weingartner01}; the distribution of grain sizes for both components follows the classical Mathis-Rumpl-Nordsieck distribution \citep{Mathis77}. Dust is distributed in a fixed toroidal structure described by six parameters:
\begin{itemize}
\item Radial extent of the torus structure, $R = R_\text{outer}/R_\text{inner}$;
\item Half-opening angle $\theta$ measured from the pole;
\item Inclination angle $i$ measured from the pole, which determines whether the sightline to the accretion disc is obscured;
\item The optical depth of the dust clouds at $9.7\mu$m, $\tau_{9.7}$;
\item Spatial distribution parameters $p$ and $q$ setting the radial and polar density distributions of the dust.
\end{itemize}
Specifically, the dust density distribution $\rho$ is given by
\begin{equation}\label{eq6}
\rho(r, \theta) \propto r^{-p} e^{-q |\cos(\theta)|},
\end{equation}
where $r,\theta$ are the usual polar cylindrical coordinates and where $R_\text{inner} = 0.5 \text{pc} \times \sqrt{L_\text{bol}/10^{11} L_\odot}$ is the sublimation radius of graphitic dust. In this work, we use the publicly-available SKIRTOR simulation, whose parameter grid spans $R\in[10,30]$, $\theta \in[10,80]$ and $i\in[0,90]$ in steps of $10$, $\tau_{9.7}\in[3,11]$ in steps of $2$, and $p, q \in[0,1.5]$ in steps of $0.5$ \citep{Stalevski16}. There are therefore $6$ physical parameters and $19,\!200$ models in total.

We require an estimated luminosity and spectral shape of the accretion disc of the quasars in order to predict the re-processed IR emission for each set of parameters. The re-processed emission is not very sensitive to the spectral energy distribution of the incident radiation, but it is very sensitive to the integrated incident flux. Therefore, the default accretion disc SED in SKIRTOR is not appropriate for us to use, since it lacks a break in the power-law at $\lambda_\text{rest}\sim4000$\AA, which is very prominent in luminous type-1 quasars. Without this point of inflection, the SED would greatly over-estimate the total luminosity in the rest-frame far-UV for a given brightness at $1\mu$m. In addition, our MRS observations cover precisely the transition between the accretion disc and the torus at $\lambda_\text{rest}\sim1\mu$m: in order to perform quantitative inference, we need to fit and/or assume the accretion disc power-law shape anyway. We therefore opt to fit the torus and accretion disc SED consistently by using complementary observations of the $z>7$ quasars obtained with NIRSpec, the details of which are described in Section~\ref{sec:nirspec}.

For the purposes of using the accretion disc SED as an input to SKIRTOR, detailed decomposition of the quasar emission into a power-law and broad emission lines is not necessary $-$ indeed, the frequency resolution of SKIRTOR does not resolve any emission lines. We therefore avoid the complex process of joint fitting of the accretion disc emission and the broad Fe~{\small{II}} continuum which is commonly performed for this wavelength range (see e.g.~\citealt{Schindler20}). Instead, we impose a power-law break at $4000$\AA \ and manually select two wavelength windows free from obvious emission features on either side of the break, using both the NIRSpec and MRS spectra. At wavelengths outside of spectral coverage, we retain the default assumptions on accretion disc SED shape from SKIRTOR. The resulting complete shape is, in the rest frame:
\begin{equation}\label{cases}
  \lambda L(\lambda) \propto \begin{cases}
    \lambda^{1.2} & \text{for $0.001\leq \lambda (\mu\text{m}) \leq 0.01$}\\
    \lambda^{0} & \text{for $0.01\leq \lambda (\mu\text{m}) \leq 0.1$}\\    
    \lambda^{\beta_1+1} & \text{for $0.1\leq \lambda (\mu\text{m}) \leq 0.4$}\\    
    \lambda^{\beta_2+1} & \text{for $0.4\leq \lambda (\mu\text{m}) \leq 5$}\\    
    \lambda^{-3} & \text{for $5\leq \lambda (\mu\text{m}) \leq 870$}\\  
    0 & \text{otherwise.}  
  \end{cases}
\end{equation}
Note that we follow SKIRTOR's assumption that the accretion disc continues at a constant slope in the IR up to $5\mu$m; this assumption makes a negligible difference to the energy balance of the system, but it is not universal (e.g.~\citealt{Collinson17}). We measure slope indices $\beta_1 = (0.64, 0.30, 0.47, 0.95)$ and $\beta_2 = (-0.55, -0.80, -0.05, -0.25)$ for the four quasars, in order of descending redshift.

The publicly-available runs of SKIRTOR are rescaled to use the accretion disc SED of each quasar using the same procedure as \citet{YangG20}. Briefly, we convert the outputs of SKIRTOR into $F(\lambda)$, normalise the measured SED to match the total luminosity of the SED assumed in the SKIRTOR model, and rescale the direct and scattered components of the torus emission by the wavelength-dependent ratio of the new and default accretion disc SED. Finally, we rescale the resulting disc and torus SEDs to match the observed luminosities of our quasars following the scaling relations of \citet{Fritz06}: $R_\text{inner} \propto \sqrt{L_\text{bol}}$ and $M_\text{dust} \propto L_\text{bol}$.
The results are modified SKIRTOR outputs containing all components of the disc and torus emission in a manner consistent with the SED measured from the NIRSpec and MRS spectra; importantly, this procedure preserves energy balance such that the light re-processed by the torus is consistent with the observed accretion disc luminosity assuming the SED from Eq.~\ref{cases}.

After subtracting the detected broad emission lines, we re-bin the MRS spectra to match the wavelength sampling of the SKIRTOR models and 
compute the $\chi^2$ metric over the range $0.7 \leq \lambda_\text{rest} (\mu\text{m}) \leq 2.0$ for quasar J1342+0928 (which is missing MRS CH4) and $0.7 \leq \lambda_\text{rest} (\mu\text{m}) \leq 2.7$ for the other quasars. We only consider observational uncertainties, as SKIRTOR outputs does not provide modelling uncertainties. We compute the likelihood of each set of parameter values as $\mathcal{L} = \exp \left(-0.5 (\chi^2 - \chi^2_\text{min}) \right)$, where $\chi^2_\text{min}$ is the minimum $\chi^2$ value corresponding to the optimal set of parameters.

\subsection{Results}

\begin{figure*}
\centering
\includegraphics[width=0.79\textwidth]{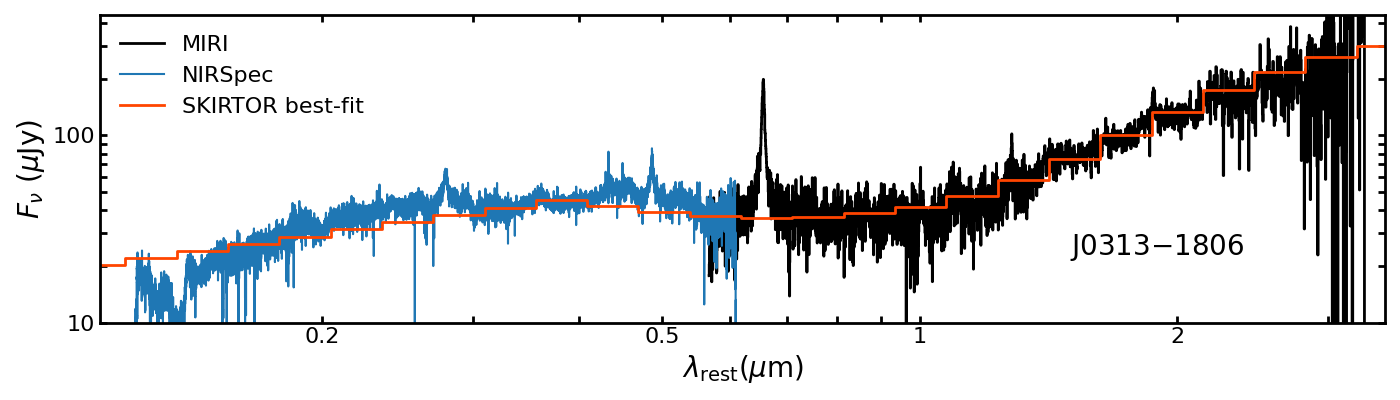}
\includegraphics[width=0.79\textwidth]{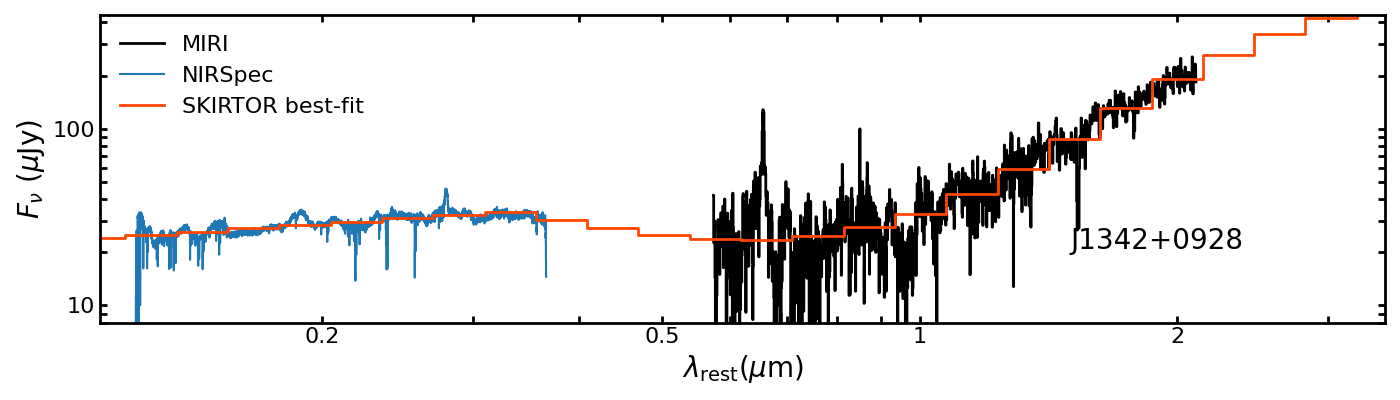}
\includegraphics[width=0.79\textwidth]{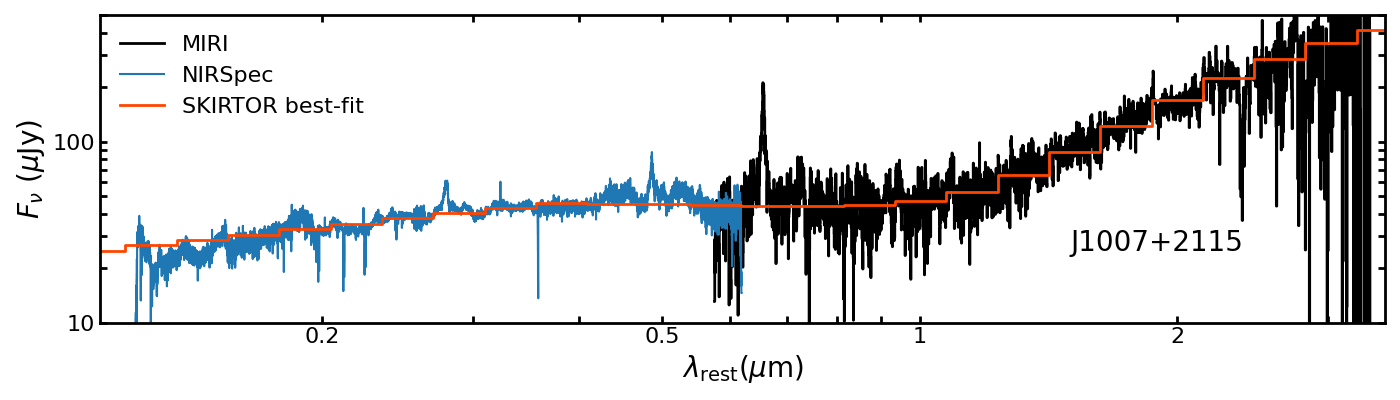}
\includegraphics[width=0.79\textwidth]{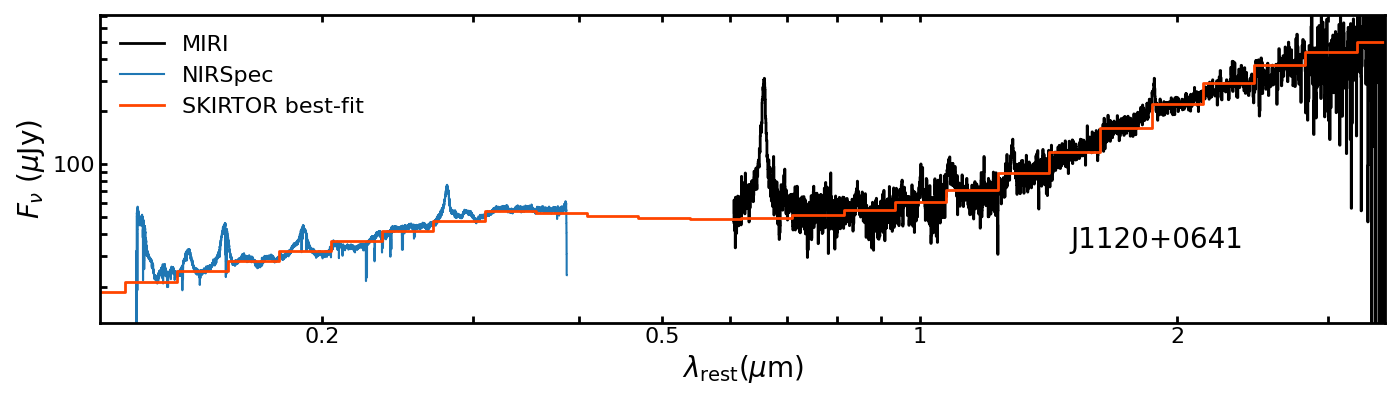}
\caption{Optimal fit to the torus emission for the four quasars (red). The NIRSpec spectra (blue) are used to estimate the SED of the accretion disc, which is then processed through SKIRTOR to predict the torus emission. The fit is conducted over the MRS spectra (black) after subtracting the detected broad emission lines. }
\label{fig:SED}
\end{figure*}

The SKIRTOR models which minimise $\chi^2$ are shown in Figure~\ref{fig:SED}. The optimal models for all four quasars are in excellent agreement with the data, with  reduced $\chi^2$ values in the range $0.85-0.95$. We show the marginalised posteriors for the $6$ physical parameters in Figure~\ref{fig:posterior} for quasar J0313$-$1806, and for the other three quasars in Appendix~\ref{appendix1}. Some of the parameters are well-constrained ($i$, $\theta$, $p$) and some poorly constrained ($\tau_{9.7}$, $R$, $q$); this split is consistent across all four quasars.

In more detail, all four quasars prefer a face-on orientation, i.e.~$i<\theta$, with a peak at $i=0\deg$ for 3/4 quasars. This is completely expected, since the accretion discs of the quasars are blue and extremely luminous. The fact that a tail to larger inclinations appears to be permitted is due to the possibility of scattering of the accretion disc light through gas in the torus without encountering a dust clump; however, the lack of attenuation of the BLRs by either dust or gas independently rules out this possibility.

The half-opening angle of the torus, $\theta$, is the best-constrained parameter. This is due to the opening angle sensitively setting the ratio between the optical and near-infrared emission of AGN (see \citealt{Stalevski16}). Indeed, for all other geometric parameters being fixed, the opening angle sets the projected cross-section of the inner torus which is visible from our line of sight, i.e.~the angular size on the sky of the hottest torus dust. All four quasars have a strong peak in the likelihood at $\theta=20\deg$, with J1120+0641 and J1007+2115 also allowing $\theta=30\deg$ at $1\sigma$. Galaxy mapping in the surroundings of type-1 luminous quasars indicate a total opening angle of the UV ionisation cone of $30-60^\circ$, corresponding to a half-opening angle of $15^\circ\lesssim\theta \lesssim 30^\circ$ \citep{Trainor13, Borisova16, Bosman20, Protusova25}. Our SKIRTOR results appear to support this picture. In contrast, studies modeling the tori of lower-$z$ fainter AGN have typically preferred ``slimmer'' tori with $40^\circ\lesssim\theta \lesssim 60^\circ$, but with large scatter \citep{Mor09,Mor12,Martinez17}. However, these studies are based on much fainter quasars; to untangle the multiple possible sources of this difference (i.~e.~the effects of redshift, luminosity, obscuration), large samples of objects with $L_\text{bol}>10^{47}$ need to be studied at later times. This will be enabled, for instance, by the \textit{SPHEREx} mission (see \ref{sec:BB}).

Finally, the power-law index of the radial density distribution, $p$, is also tightly constrained; all four quasars prefer the highest available value of $p=1.5$, which corresponds to the model where dust is maximally centrally concentrated (see Eq.~\ref{eq6}). Models employing SKIRTOR for SED-fitting of AGN and galaxies, such as X-CIGALE, have assumed a slightly shallower $p=1.0$ \citep{XCIGALE}. This choice is motivated by work on the IR emission variability of low-$z$ AGN with bolometric luminosities $100$ times lower than our objects \citep{Mor09, Kaspi99}, but it sufficient for the purposes of most SED fitting since observed broad-band photometry in the NIR would rarely be capable of constraining $p$ and $q$ unless the AGN are face-on and the host galaxy contribution to the rest-frame optical is negligible \citep{Stalevski12}. In contrast, our observations are very sensitive to $p$, as it sets the shape of the transition from accretion disc to torus at $\lambda_\text{rest}\sim1\mu$m by effectively changing the fraction of the dust with $T>1000$ K. The fact that our quasars prefer the steepest available $p$ may imply that in the future we will require SKIRTOR models with values larger than $p=1.5$. Indeed, some (fainter) AGN tori modelled with clumpy+diffuse dust \citep{Mor12} as well as purely clumpy dust \citep{Martinez17} prefer even higher values, $p=2$. 
Regardless, we are satisfied with our optimal fits since they display excellent reduced $\chi^2$ values without going beyond the bounds of classical assumptions.

Other physical parameters are less well constrained. The lowest value of dust optical depth, $\tau_{9.7}=3$, is strongly excluded for all quasars, while $\tau_{9.7}\geq 7$ is required for 3/4 quasars. Since $\tau_{9.7}$ is strongly tied to total dust density and therefore mass, the rejection of low $\tau_{9.7}$ values implies that the observed NIR luminosities would not be possible if the mass of hot dust in the torus were too low. The parameters $q$ and $R$ mostly affect the visibility and strength of the silicate emission feature and longer-wavelength dust continuum from the torus, and are therefore poorly constrained for our available spectroscopic coverage.

In conclusion, we find that the SKIRTOR model is an excellent quantitative description of the observed rest-frame optical and NIR SEDs of our quasars. The rest-frame wavelength range used in model fitting, $0.7\leq\lambda_\text{rest} (\mu\text{m})\leq2.7$, contains sufficient information to constrain the inclination $i$, half-opening angle $\theta$, and radial dust index $p$; the constraints on these parameters are in agreement with classical assumptions about luminous quasars. In Section~\ref{sec:dustmass} we will discuss the dusty torus masses implied by our best-fit models.

\subsection{Redshift evolution?}

\begin{figure*}
\centering
\includegraphics[width=0.79\textwidth]{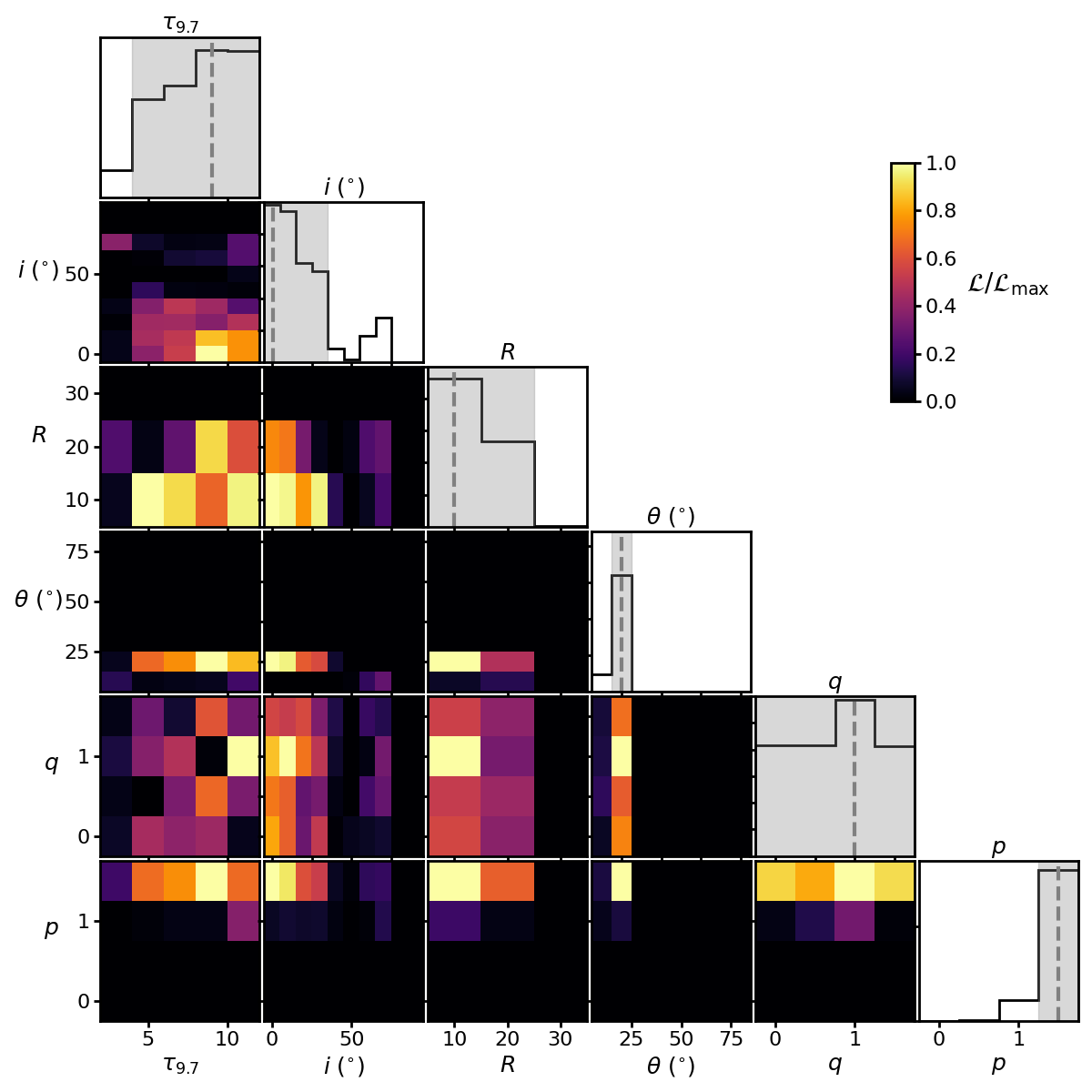}
\caption{Posterior likelihood of torus parameters for quasar J0313$-$1806. The marginalised likelihoods show good constraining power on several parameters (opening angle $\theta$, inclination $i$, radial dust distribution index $p$) but not on others (polar dust distribution index $q$, optical depth $\tau_{9.7}$, radial extent $R$). The grey dashed lines and shaded regions show the parameter values with the highest marginalised likelihoods and $1\sigma$ confidence intervals.}
\label{fig:posterior}
\end{figure*}

Detailed modelling of torus emission for unobscured quasars of similar luminosities and BH masses at $z<5$ has seldom been conducted, largely due to the sparsity of suitable spectroscopic observations. Instead, in order to assess whether hot dust properties evolve across redshift, we must either rely on purely observational characteristics (the IR excess) or mimic the modelling employed by past studies. In the following subsections, we do both.

\subsubsection{The hot torus as black-body emission}\label{sec:BB}

As noted near the beginning of Section~\ref{sec:torus}, by far the most widespread modelling assumption for quasar SEDs in the NIR is that of a power-law continuum originating from the accretion disk together with a single-temperature black-body component corresponding to the hottest dust inside of the torus \citep{Leipski14,Hernan16,Bosman24}. To compare with previous studies, we implement such a model with \texttt{Sculptor}. We vary four parameters, corresponding to the accretion disc power-law index and its normalisation, and black-body emission temperature and its  normalisation. Similar to our SKIRTOR fits, we subtract the broad emission lines before fitting over $0.7\leq\lambda_\text{rest} (\mu\text{m})\leq2.7$. We also repeat the procedure using a modified black-body with $\beta=2$.

\begin{figure}[t]
\includegraphics[width=\columnwidth]{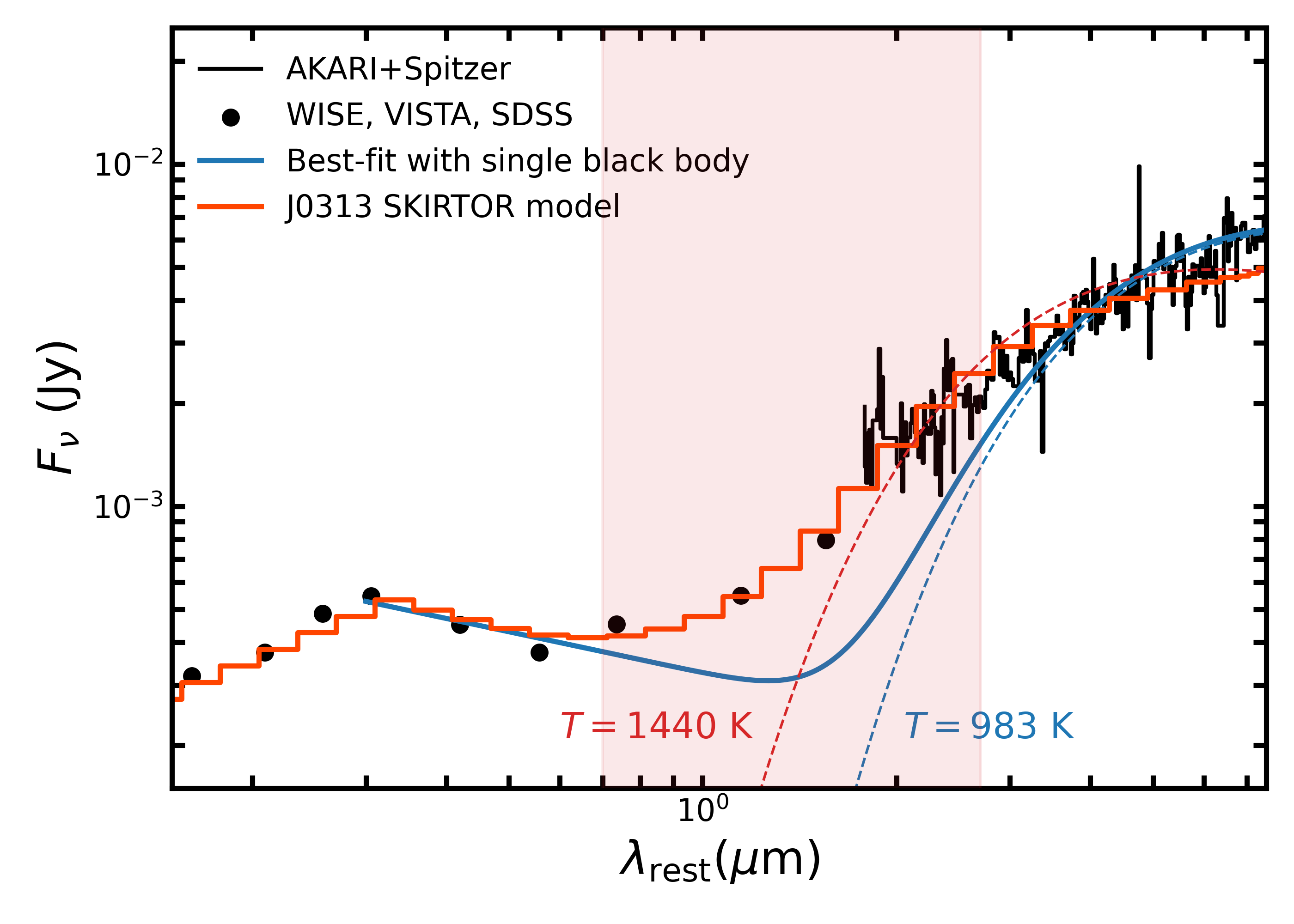}
\caption{Photometry and spectroscopy of quasar SBS 1040+567 at $z=1.96$, compared with multiple models. Using the data in black, the best-fit power-law plus accretion disc fit return a very cold hot torus temperature of $T=983$ K \citep{Hernan16}. However, a direct rescaling of our SKIRTOR model for the $z=7.5$ quasar J0313$-$1806 (red) provides an equivalent or better fit, despite the fact the hot dust temperature we would measure for this fit in the corresponding wavelength range is $T=1440$ K. This temperature bias likely comes from the fact that the torus contains dust at multiple temperatures which dominate at different rest-frame wavelengths. The $\lambda_\text{rest}$ range covered by our MRS spectra is shaded in red.}
\label{fig:lowz}
\end{figure}

We find that a single-temperature black-body (whether modified or not) is sufficient to capture the shape of the continuum in the MRS spectra (reduced $0.9<\chi^2<1.2$), as we explored in more detail for J1120+0641 in \citet{Bosman24}. The resulting hot dust temperatures are in the range $1400\lesssim T(\text{K}) \lesssim 1515$ for all four quasars assuming classical (optically thick) black-body emission, and $T>2300$ K for all four quasars assuming $\beta=2$ modified (optically thin) black-body emission. The latter value is likely unphysical as a single temperature for the entire hot torus, since graphite dust will survive $<1000$ yr at temperatures $T>2100$ K \citep{Phinney89}.

The dust temperatures assuming black-body emission from a single dust component are systematically higher than observed in similar quasars at $z<3$: $T>1400$ K for all four $z>7$ quasars, compared to only $1$ quasar out of $85$ at $z<3$ \citep{Hernan16}. However, upon closer comparison, we realised that such a shift is expected purely as a result of systematic bias in the sampling of rest-frame wavelengths at $z>7$ compared to $z<3$. Low-$z$ observations have employed spectroscopy from \textit{Spitzer} \citep{Spitzer} and sometimes AKARI \citep{AKARI} which cover rest-frame wavelengths $2\lesssim\lambda_\text{rest} (\mu\text{m})\lesssim10$ at $z<3$, where dust with a temperature $700<T(\text{K})<1200$ dominates. In contrast, at $0.7\lesssim\lambda_\text{rest} (\mu\text{m})\lesssim2.7$ as covered by our $z>7$ MRS spectra, dust with $T>1000$ K dominates and dust with $T=700$ K would not be detectable at all. Therefore, if real dusty tori contain dust at a continuum of temperatures (as modelled for example in SKIRTOR), then the assumption of a single temperature would bias measurements depending on which rest-frame wavelengths are most sampled and drive the fit.

We show an illustration of the temperature bias in Figure~\ref{fig:lowz}. We selected quasar SBS 1040+567 at $z=1.96$, for which \citet{Hernan16} measured one of the ``coldest'' hot torus temperatures in their sample of $85$ objects: $983$ K. The optimal power-law plus black-body fit from that study is plotted in blue, showing that the fit is optimised to reproduce the highly spectroscopically sampled \textit{Spitzer} data at $\lambda_\text{rest} > 2\mu$m but under-estimates the WISE photometry at $\lambda_\text{rest}\sim1.5\mu$m in doing so. In contrast, this is exactly the range that our measurement is driven by (shaded area). In red, we show the best-fit SKIRTOR physical torus model for quasar J0313$-$1806, rescaled to the accretion disc shape and luminosity of SBS 1040+567. The SKIRTOR model provides a much better description of the observed WISE photometry SBS 1040+567 without a significantly worse agreement with the \textit{Spitzer} data, despite the fact it would correspond to a temperature of $T>1400$ K if measured over the shaded region. 

We interpret this bias as strong further confirmation that the torus contains dust at more than a single temperature. Any meaningful comparison across redshift therefore requires both a more complex model and equivalent sampling of rest-frame wavelengths; the latter is nearly impossible possible with current \textit{Spitzer} and AKARI observations \citep{Jun15}. However, the ongoing \textit{SPHEREx} mission is obtaining finely-sampled photometry of over 1.5 million quasars at $0<z<7$ over the $0.75<\lambda_\text{obs} (\mu\text{m})<5.0$ wavelength range. Rather than dealing with complex observational and systematic uncertainties in previous data, we will present a large-scale comparison of physically-motivated  torus properties based on \textit{SPHEREx} in upcoming work.

\subsubsection{The $2\mu$m IR excess}\label{sec:excess}

In order to assess potential evolution in NIR properties in a model-independent way, we design a metric to capture the amount of ``upturn'' in the quasar continuum between the optical wavelengths ($0.6-0.8\mu$m) which should be dominated by the accretion disc, and the NIR wavelenths ($2\mu$m) which are expected to show extra emission from the torus. We define the $2\mu$m IR excess as follows: first, we fit a power-law using the observed flux of the continuum at $\lambda_\text{rest}\sim 0.6\mu$m and $\sim 0.8\mu$m, which bracket the H$\alpha$ emission line. We then extrapolate this power-law to $\lambda_\text{rest}=2\mu$m to estimate the expected brightness of the continuum in the absence of a torus. The $2\mu$m IR excess is then the ratio of the observed flux at $\lambda_\text{rest}=2\mu$m to the torus-free expectation.

Our definition has the advantage of being insensitive to wavelength sampling since it only requires $3$ values of flux: one at $\lambda_\text{rest}=2\mu$m and two measurements on either side of the H$\alpha$ line. Note that this metric is different from the ``$2.3 \mu$m flux density ratio'' of \citet{Jun13}, who use low ratios in $\lambda F_\lambda$ between $2.3\mu$m and $0.5\mu$m to identify dust-poor quasars. 
A low raw flux ratio, without subtracting the expected luminosity of the accretion disc, can only be interpreted as a lack of dust emission in the context of the accretion disc SED, while our method has the advantage that a $2\mu$m IR excess consistent with a value of $1$ indicates dust-poorness regardless of the accretion disc SED. 

We measure the $2\mu$m IR excess in our four quasars as well as in a sub-set of $45$ quasars from \citet{Hernan16} for which suitable observations exist. For example, SBS 1040+567, shown in Fig.~\ref{fig:lowz}, is not usable since its VISTA Hemisphere Survey \citep{VISTA} K-band photometry encompasses the H$\alpha$ emission line; flux measurements on either ide of the lines are therefore not possible. For a fair comparison, we re-measure the optical slopes of our quasars after transposing them to $z=1.5$ (the median of the low-$z$ sample) and binning their spectra to a resolution corresponding to VISTA and WISE photometric bands. The resulting slopes are slightly redder than we inferred using the full NIRSpec and MRS spectra in Section~\ref{model}, presumably because the torus is already contributing some emission for photometric observations centred at $\sim 0.8\mu$m. Nevertheless, this definition allows us to retain the largest number of low-$z$ quasars for comparison. The $2\mu$m IR excess is significantly correlated with optical slope: for an equal bolometric luminosity and identical emission from a dusty torus, quasars which are bluer in the optical will display a larger excess. We estimate uncertainties by perturbing the photometry of the low-$z$ quasars by their observational uncertainties. 

\begin{figure}[t]
\includegraphics[width=\columnwidth]{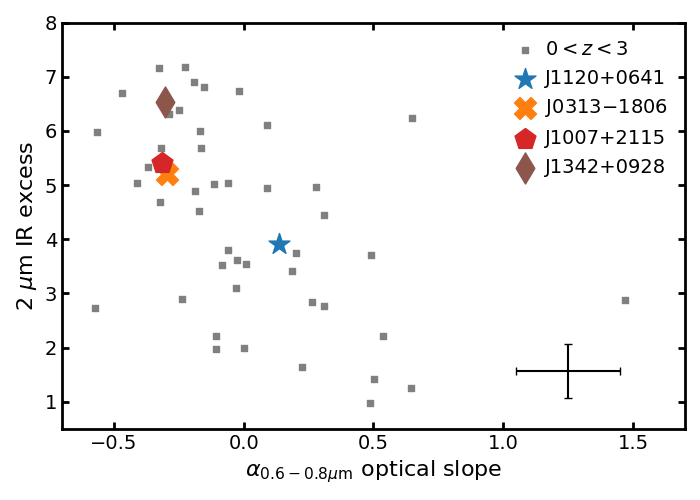}
\caption{The $2\mu$m IR excess plotted against optical ($0.6-0.8\mu$m) slope, for our $z>7$ quasars and $45$ comparable quasars at $0<z<3$ from \citet{Hernan16}. We define the $2\mu$m IR excess as the ratio of the measured flux at $2\mu$m to the torus-free expectation from extrapolating the accretion disc alone. No statistically significant differences between the low-$z$ and high-$z$ samples are present. Representative uncertainties are shown in the bottom right.}
\label{fig:excess}
\end{figure}

We show the result in Figure \ref{fig:excess}. Our quasars have $2\mu$m IR excesses between $4$ and $7$, meaning that their brightness at $2\mu$m is $4-7$ times higher than expected from extrapolating emission from the accretion disc. At $z<3$, a broad range of $1-7.5$ is seen, which encompasses our objects. The low-$z$ (high-$z$) sample has a mean of $4.33$ ($5.26$) and a standard deviation of $1.77$ ($0.93$). At fixed optical slope, the low-$z$ quasars display significant scatter in their $2\mu$m IR excess, implying that there exists variety in their torus properties, which we will explore in future work. Our high-$z$ quasars sit roughly in the median of this scatter.  
We find no signs of redshift evolution in the $2\mu$m IR excess from a two-sample KS test ($p=0.29$). The accretion disc power-law slopes in the optical similarly show no evidence of a difference between the samples ($p=0.12$). 
We note that $3/45$ quasars are consistent with the absence of a torus at $z<3$ ($2\mu$m IR excess consistent with $1$). Consistent with \citet{Jun13}, we find that these objects all have slight red accretion disc SEDs. Given their abundance, we would need to observe $>15$ quasars with at $z>7$ to be able to constrain the existence of such objects at early times. Much like the more detailed torus modelling, the study of the $2\mu$m IR excess in large samples at $z<7$ will be tremendously facilitated by the \textit{SPHEREx} mission.

\section{Discussion}

\subsection{Mass of the torus and quasar lifetime} \label{sec:dustmass}

Using the best-fit SKIRTOR models for our MRS spectra, we can estimate the total dust mass (of all temperatures) contained in the torus. The SKIRTOR outputs include information on the total mass of dust for each combination of geometric parameters \citep{Stalevski16}, which we rescale to the observed bolometric luminosities of our quasars using the scaling relations of \citet{Fritz06}. Despite the fact that not all parameters of the model are well constrained, all allowed models for all four quasars have dust masses in the range $M_\text{dust} = (1-4) \cdot 10^6 M_\odot$. The largest uncertainty comes from the physical extent of the torus, $R$: while all four quasars prefer $R=10$, $R=20$ is still allowed at $1\sigma$, which increases the total dust mass by a factor of $2.5$. Since the dust at large distances from the accretion disc is increasingly cold, the only way to further constrain the outer geometry of the torus will be with longer far-IR wavelengths, as planned for example by the PRIMA mission \citep{PRIMA}.

For comparison, the total dust masses of the host galaxies of our four $z>7$ quasars have been estimated to be roughly $M_\text{dust} = (0.6-4.3) \cdot 10^8 M_\odot$ depending on assumptions on the dust composition and temperature \citep{Venemans17,Venemans17b,Yang20,Wang21}. The fraction of the galaxy's dust located in the accretion torus is therefore roughly $(0.2 - 7)\%$. Assuming a gas-to-dust ratio in the torus of $70-100$, and assuming a luminous efficiency of the BH accretion of $\epsilon=0.1$, the entire current matter content of the torus would be depleted in $(3-8)$ Myr if the quasars continue accreting at their observed rates. This characteristic depletion timescale is consistent with measurements of the typical time since the onset of the current bright phase for large populations of quasars at $z>6$, of the order of $\sim 1$ Myr \citep{Eilers20,Eilers21}. It therefore appears as if our quasars have been observed at a typical time within their current bright phase, with enough matter stored in their tori to maintain their luminosities for another $\sim5$ Myr (assuming direct transfer of matter from the torus to the accretion disc, with no kinetic feedback).

\subsection{Reliability of single-epoch BH mass measurements}\label{sec:reliable}

In this paper, we have explored many observational properties of $z>7$ quasars which are uniquely accessible in the rest-frame infrared. The H$\alpha$ emission line is one of the most reliable single-epoch BH mass estimators since its width and luminosity are known to correlate tightly with those of H$\beta$, which is in turn used directly for BH mass estimation through reverberation mapping in $z<3$ luminous type-1 quasars. We have found that these more accurate H$\alpha$-based masses agree fully, within expected intrinsic scatter, with tracers based on emission lines at longer wavelengths (Pa$\alpha$, Pa$\beta$) and ground-based measurements based on Mg~{\small{II}} (and to a lesser extent, even C~{\small{IV}}). There is therefore no evidence of bias in the BH mass measurements due to extinction. The ratios of the H$\alpha$, Pa$\alpha$ and Pa$\beta$ emission lines deviate from expectations for both case A and case B recombination, but they do so in the same direction as $0<z<4$ quasars, and show no sign of BLR extinction. In short, we have failed to find any signs of physical differences in $z>7$ quasars which would render single-epoch BH mass estimators biased or unexpectedly inaccurate compared to $0<z<4$ quasars.

Recently, \citet{King24} has suggested that the masses of bright quasars may be over-estimated by large factors if their observed accretion disc luminosities have been subject to beaming by fast disc-driven outflows. Beaming, which occurs in e.g.~ultra-luminous X-ray sources, causes the luminosity of the accretion disc to be highly non-isotropic $-$ much brighter in the polar direction $-$ in the presence of large mass outflow rates from the disc \citep{King01, King23}. This can lead to over-estimation of the bolometric luminosity by a factor up to $\sim70$. The observed IR luminosities of our $z>7$ quasars offer an elegant refutation of this possibility: in order to power the observed luminosity from the dusty torus, our modeling shows that a dust mass of at least $10^6 M_\odot$ is required to re-process the assumed (isotropic) bolometric luminosity of the accretion disc. In other words, the torus ``sees'' the same luminosity being emitted by the accretion disc as we do; throughout our modelling we have preserved energy balance. In contrast, if beaming were present in an amount sufficient to bias the BH mass measurement (i.e.~factors $>10$), then it would boost the accretion disc and broad line luminosities towards the observer, but not towards the torus $-$ 
the amount of dust required to reproduce the observed IR luminosities would need to increase by the square of the factor by which the BH mass is over-estimated, which becomes comparable to the mass of the whole host galaxy.
\footnote{The second argument in \citet{King24}, that broad lines are broadened primarily by outflows, does not apply to our measurements since the lines we use are not blueshifted compared to the host galaxies. In broad lines which are, indeed, broadened by outflows (C~{{IV}}, Si~{{IV}}; see e.g.~\citealt{Onoue20}) these outflows are always asymmetric.}


\subsection{Assembly of the first SMBHs}\label{sec:growth}

\begin{figure}[t]
\includegraphics[width=\columnwidth]{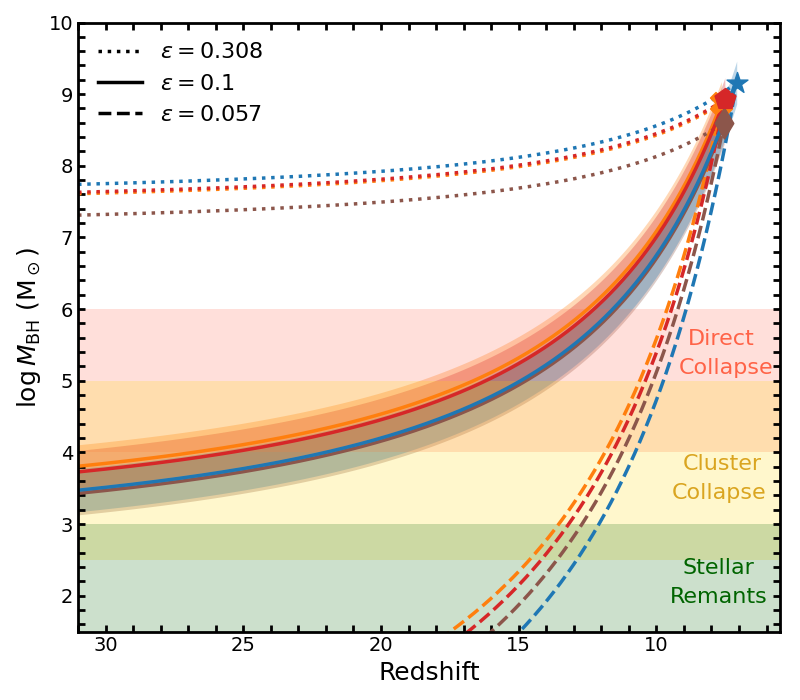}
\caption{Trace-back of the growth of SMBHs in $z>7$ quasars to $z=30$ under various assumption for the radiative efficiency $\epsilon$. Under the classical assumption of $\epsilon=0.1$, the first SMBHs cannot have grown from stellar remnants from Pop III stars even when accreting continuously at the maximal rate. For a lower radiative efficiency, such as the one expected for spin-free BHs, this remains possible. The widths of the $\epsilon=0.1$ tracks reflect systematic uncertainties on the BH mass measurements.}
\label{fig:growth}
\end{figure}

Using our updated BH masses, we revisit the possible assembly histories of $z>7$ SMBHs, as shown in Figure~\ref{fig:growth}. For the purposes of this discussion, we consider only the systematic uncertainties on the BH masses, since they dominate over measurement uncertainties by a factor of $\sim10$. We also neglect type-2 AGN at $z>7$ (LRDs and X-ray sources) since the systematic uncertainties on their inferred BH masses are not known quantitatively. 

If the growth of $z>7$ SMBHs is limited to the classical assumption of accretion with radiative efficiency of $\epsilon = 0.1$ (meaning that $10\%$ of accreted mass is converted into luminosity), then the limit of the Eddington luminosity would require that they grow from ``seed'' BHs at $z=30$ with masses at least $M_\text{seed}\gtrsim 10^4 M_\odot$. Such seed BHs are not compatible with expectations for remnants of the first generation of stars (Pop III), which should not exceed $M_\text{seed} = 1000 M_\odot$ (e.g.~\citealt{Madau01}). Formation via dynamical cluster collapse is marginally allowed, as it could produce $M_\text{seed} \leq 10^4 M_\odot$\footnote{We differ on this point from \citet{Wang21} due to the difference in assumed cosmology: $(h,\Omega_M) = (0.6774, 0.3089)$ in this paper compared to $(h,\Omega_M) = (0.7, 0.3)$ in \citet{Wang21}.} (e.g.~\citealt{Portegies04, Rantala25, Reinoso25, Vergara25}). Even though J1120+0641 hosts the most massive known BH at $z>7$, the most stringent constraints on BH formation come instead from J0313$-$1806 due to its higher redshift $z=7.642$, despite the fact that its BH is inferred to be smaller by a factor of $\sim2$. 

However, a radiative efficiency of $\epsilon=0.1$ is not expected to be universal nor constant. As noted already by \citet{Thorne74}, the radiative efficiency of an accreting BH may be at least as low as $\epsilon=0.057$ for a black hole with zero spin, and at least as high at $\epsilon=0.308$ if the accretion disc co-rotates with the BH (see also \citealt{Shapiro05, Sadowski09}, and e.g.~\citealt{Ricarte23} for BH spin-down via jet launching). A change in $\epsilon$ would not affect the accuracy of the single-epoch mass estimators, since the BLR and accretion disc would brighten (or dim) proportionately with each other as long as the accretion remains sub-Eddington\footnote{At most, the masses would be biased by the ratio of the true $\epsilon$ to $0.1$, and would still be contained within uncertainties.}. However, the possible assembly histories would be dramatically modified: for $\epsilon=0.308$, much more mass is ``radiated away'', requiring BHs with masses $M_\text{seed}\gtrsim 10^7 M_\odot$ at $z=30$, which could not have formed by any known physical process, even direct collapse of pristine gas clouds (e.g.~\citealt{Begelman06}). A bias towards high $\epsilon$ might be expected in our sample, since quasars at $z>7$ are selected in full-sky surveys to be the brightest objects: for constant $M_\text{BH}$ and $\dot{M}$, quasars with a higher $\epsilon$ would be brighter and therefore easier to discover. 

Conversely, $\epsilon<0.1$ could also be argued to be more likely for $z>7$, since chaotic morphologies and turbulent feedback in early quasar host galaxies are less likely to allow accretion discs to always co-rotate with the SMBH. Low $\epsilon$ has also been theorised to arise from change in accretion morphology at high inflow rates (e.g.~\citealt{Madau14}), although it would then occur also in some AGN at late times. As shown in Figure~\ref{fig:growth}, spending even a short amount of time ($\sim 350$ Myr) in a low-$\epsilon$ phase would enable all of our BHs to easily grow from Pop III stellar remnants. The time spent accreting could be even lower if the accretion disk counter-rotates compared to BH spin, in which case the efficiency can be as low as $\epsilon=0.037$ \citep{Thorne74}. By using their ionised proximity zones as ``total photons counters'', \citet{Davies19} found strong evidence for a lifetime-integrated $\epsilon<0.1$ in quasars J1342+0928 and J1120+0641 from our sample, preferring an extremely low $\epsilon\sim 0.001$. However, even a much more modest change would be sufficient to allow $z>7$ SMBHs to originate from Pop III remnant BHs: for $\epsilon=0.08$, just $20\%$ below the canonical value, all of our quasars could grow from $100-500M_\odot$ Pop III remnants at $z=30$.  A simple solution to the ``SMBH growth problem'' would therefore be that $\epsilon$ varies over the lifetime of early SMBHs, with at least one phase of low radiative efficiency due to low BH spin or accretion counter-rotating with the central BH, and maybe a phase of higher efficiency near the time of observation to boost detectability (see also \citealt{King06}). 

\section{Summary}\label{sec:ccl}

In this study, we have used spectroscopic data from the MIRI-MRS and NIRSpec instruments onboard \textit{JWST} to observe and model the rest-frame IR emission from the four highest-redshift currently known quasars. We detect the broad H$\alpha$ emission line and an upturn in the continuum emission at $\lambda_\text{rest}\gtrsim1\mu$m in all four quasars, and detect the Pa$\alpha$ and Pa$\beta$ broad emission lines in three of the quasars.

We find that the BH masses implied by the FWHMs and luminosities of all broad lines are consistent with each other within expected systematic scatter, and they are consistent with previous ground-based BH mass estimates from the Mg~{\small{II}} broad emission line. We update the BH masses of the objects using the single-epoch mass estimator based on the H$\alpha$ emission line, which has the smallest systematic uncertainties compared to reverberation mapping measurements at late times. We find that quasar J1120+0641 hosts the most massive known SMBH at $z>7$, with $M_\text{BH} = 14.5\pm 1.2 \cdot 10^8 M_\odot$ within $0.3$ dex systematic uncertainties. 

The flux ratios of the H$\alpha$, Pa$\alpha$ and Pa$\beta$ emission lines are not consistent with case A nor case B recombination, in the same manner as seen in nearly all $z<3$ quasars. The deviations from case A/B recombination rate are not consistent with extinction of the BLR, but rather with high densities and ionisation intensities from a central source, as we demonstrate using a simple \texttt{Cloudy}-based model. Quasar J0313$-$1806 displays an unusually weak Pa$\alpha$ emission line compared to Pa$\beta$ and H$\alpha$, which may be related to a high ionisation parameter of its accretion disc, and may be connected to its ability to drive the extreme broad absorption line outflows seen in the object (up to $0.18c$). At the moment, no quasars with similarly extreme masses and outflows exist at $z<3$ with observations of all three emission lines; any connection is purely speculative.

We confirm previous findings that none of these SMBHs could have grown from the remnants of Pop III stars if the radiative efficiency of accretion has the constant canonical value of $\epsilon = 0.1$. Quasar J0313$-$1806, with $\log M_\text{BH} = 7.52\pm 0.66 M_\odot$, is at the moment the SMBH whose early growth is the most difficult to explain, given its higher $z=7.642$. However, with even a modestly lower $\epsilon=0.08$, all four quasars can grow from BH seeds in the range $100-500 M_\odot$ at $z=30$ assuming continuous accretion. With an even lower $\epsilon=0.057$, as expected for black holes with zero spin, the BHs would only need to accrete continuously at the zero-spin efficiency for $\sim350$ Myr (i.e.~since $z=12$) to reach their observed masses. 

Turning to the continuum IR emission, the upturn at $\lambda_\text{rest} = 1\mu$m is an unambiguous detection of emission from a dusty torus in all four quasars. We model the torus using the numerical model SKIRTOR, in which the structure is toroidal and described by $6$ parameters, which we sample using a total of 19,\!200 models. We use NIRSpec observations of the quasars to measure the SED shape of their accretion discs in the rest-frame UV and optical wavelengths, which we define as a broken power-law with a break at $4000$\AA. The SKIRTOR models are calibrated to the accretion disc SED of each quasar, ensuring energy balance between the observed accretion disc and the IR continuum re-processed by the dusty torus. 

The $0.7<\lambda_\text{rest} (\mu\text{m}) <2.7$ range is very well fit by SKIRTOR with reduced $\chi^2$ in the range $0.85-0.95$. We find that $3$ torus parameters are consistently constrained in all four objects. The inclination of the torus is face-on, as expected for luminous type-1 quasars with no detectable obscuration of the BLR. The half-opening angle of all quasars is in the range $20^\circ-30^\circ$, consistent with classical expectations of the opening angle of the ionisation cone. Finally, all quasars prefer the radial profile of the dust density distribution to be maximally centrally-concentrated ($p=1.5$). While this may imply that more extreme models are required, steeper dust gradients have not been invoked in the literature.

The optimal SKIRTOR models imply total dust masses in the torus of $(1-4) \cdot 10^6 M_\odot$ for all quasars, corresponding to $0.2-7\%$ of the total dust mass estimated from sub-mm observations of their host galaxies. Assuming a gas-to-dust ratio in the range $70-100$, the SMBHs would consume the entire matter content of the torus structure in $3-8$ Myr at their current rates of accretion. This lifetime agrees with order-of-magnitude estimates of the time since the onset of the current luminous phase in $z>6$ quasars, $\sim1$ Myr.

We find that using a black-body of a single temperature to describe the hottest component of the torus is sufficient to explain the continuum upturn in our MRS observations (reduced $\chi^2$ range $0.9-1.2$) but that it creates a bias compared to observations of similarly-bright quasars at $z<3$. Since real tori are not composed of dust of a single temperature, a higher sampling of longer rest-frame wavelengths leads to colder inferred temperatures, and vice-versa. We demonstrate this by using a $z=2$ quasar for which a cold temperature $T=983$ K would be inferred using \textit{Spitzer} data, but which is fit equally well by our optimal SKIRTOR solution for the $z=7.6$ quasar which has an equivalent single-component dust temperature of $T=1440$ K. Instead, we use the empirical $2\mu$m IR upturn to show that there are no differences between the optical and near-IR SEDs of $z>3$ and $z>7$ quasars.

In summary, we have found no indications either in the rest-frame IR broad emission lines nor in the torus SED of any peculiarities of $z>7$ quasars compared to similarly-bright objects at $z<3$. Unless inferred BH masses are incorrect also at late times, it does appear that the first SMBHs, and their accretion structures, assemble in an unexpectedly short time in the early Universe.

\begin{acknowledgements}
SEIB, KP and BS are supported by the Deutsche Forschungsgemeinschaft (DFG) under Emmy Noether grant number BO 5771/1-1. AAH is supported by grant PID2021-124665NB-I00  funded by MCIN/AEI/10.13039/501100011033 and by ERDF A way of making Europe. D.L.~was supported by a VILLUM FONDEN Investigator grant (project number 16599). C.M.~acknowledges support from Fondecyt Iniciacion grant 11240336 and the ANID BASAL project FB210003. L.C.~and J.A-M.~acknowledge support by grants PIB2021-127718NB-I00 and PID2024-158856NA-I00. JFH acknowledges support from the European Research Council (ERC) under the European Union’s Horizon 2020 research and innovation program (grant agreement no.~885301) and from the National Science Foundation under grant no.~2307180. F.W.~and X.F.~acknowledge support from the US NSF Grant AST-2308258.  J. Y.~and X.F.~acknowledge support by NSF grants AST 19-08284. : G.\"{O}.~acknowledges funding from the Swedish National Space Administration (SNSA).  J.P.P.~acknowledges financial support from the UK Science and Technology Facilities Council, and the UK Space Agency. A.A.-H.~acknowledges financial support from grant PID2021-124665NB-I00 funded by MCIN/AEI/10.13039/501100011033 and by ``ERDF A way of making Europe''.  C.M.~acknowledges support from Fondecyt Iniciacion grant 11240336 and the ANID BASAL project FB210003.

In addition to the codebases mentioned in the main text, this work made use of \texttt{numpy} \citep{numpy}, \texttt{astropy} \citep{astropy}, \texttt{matplotlib} and  \citep{matplotlib} and \texttt{ipython} \citep{ipython}.

The VISTA Hemisphere Survey data products served at Astro Data Lab are based on observations collected at the European Organisation for Astronomical Research in the Southern Hemisphere under ESO programme 179.A-2010, and/or data products created thereof. 
This work is based on observations made with the NASA/ESA/CSA James Webb Space Telescope. The data were obtained from the Mikulski Archive for Space Telescopes at the Space Telescope Science Institute, which is operated by the Association of Universities for Research in Astronomy, Inc., under NASA contract NAS 5-03127 for JWST; and from the European JWST archive (eJWST) operated by the ESDC.

This work is based on observations made with the NASA/ESA/CSA James Webb Space Telescope. The work presented is the effort of the entire MIRI team and the enthusiasm within the MIRI partnership is a significant factor in its success. The 
following National and International Funding Agencies funded and supported 
the MIRI development: NASA; ESA; Belgian Science Policy Office (BELSPO); 
Centre Nationale d’Etudes Spatiales (CNES); Danish National Space Centre; 
Deutsches Zentrum fur Luftund Raumfahrt (DLR); Enterprise Ireland; 
Ministerio De Economia y Competividad; Netherlands Research School for Astronomy (NOVA); Netherlands Organisation for Scientific Research (NWO); Science and Technology Facilities Council; Swiss Space Office; Swedish National 
Space Agency (SNSA); and UK Space Agency. MIRI drew on the scientific 
and technical expertise of the following organizations: Ames Research Center, USA; Airbus Defence and Space, UK; CEAIrfu, Saclay, France; Centre Spatial de Liège, Belgium; Consejo Superior de Investigaciones Cientficas, Spain; Carl Zeiss Optronics, Germany; Chalmers University of Technology, Sweden; Danish Space Research Institute, Denmark; Dublin Institute for Advanced Studies, Ireland; European Space Agency, Netherlands; ETCA, Belgium; ETH Zurich, Switzerland; Goddard Space Flight Center, USA; Institute d’Astrophysique Spatiale, France; Instituto Nacional de Técnica Aeroespacial,Spain; Institute for Astronomy, Edinburgh, UK; Jet Propulsion Laboratory, 
USA; Laboratoire d’Astrophysique de Marseille (LAM), France; Leiden University, Netherlands; Lockheed Advanced Technology Center (USA); NOVA Opt-IR group at Dwingeloo, Netherlands; Northrop Grumman, USA; Max Planck 
Institut f\"{u}r Astronomie (MPIA), Heidelberg, Germany; Laboratoire d'Etudes 
Spatiales et d’Instrumentation en Astrophysique (LESIA), France; Paul Scherrer Institut, Switzerland; Raytheon Vision Systems, USA; RUAG Aerospace,
Switzerland; Rutherford Appleton Laboratory (RAL Space), UK; Space Telescope Science Institute, USA; Stockholm University, Sweden; Toegepast- Natuurwetenschappelijk Onderzoek (TNOTPD), Netherlands; UK Astronomy Technology Centre, UK; University College London, UK; University of Amsterdam, Netherlands; University of Arizona, USA; University of Cardiff , UK; 
University of Cologne, Germany; University of Ghent; University of Groningen, Netherlands; University of Leicester, UK; University of Leuven, Belgium; Utah State University, USA. 

All raw and calibration data used in this work from both MIRI and NIRSpec may be obtained from the online Mikulshi Archive for Space Telescopes (MAST) by using the PIDs listed in Table~\ref{tab1}.
\end{acknowledgements}

\bibliography{bibliography}{}
\bibliographystyle{aasjournal}

\appendix

\section{SKIRTOR posteriors for all quasars}\label{appendix1}

\begin{figure*}
\centering
\includegraphics[width=0.79\textwidth]{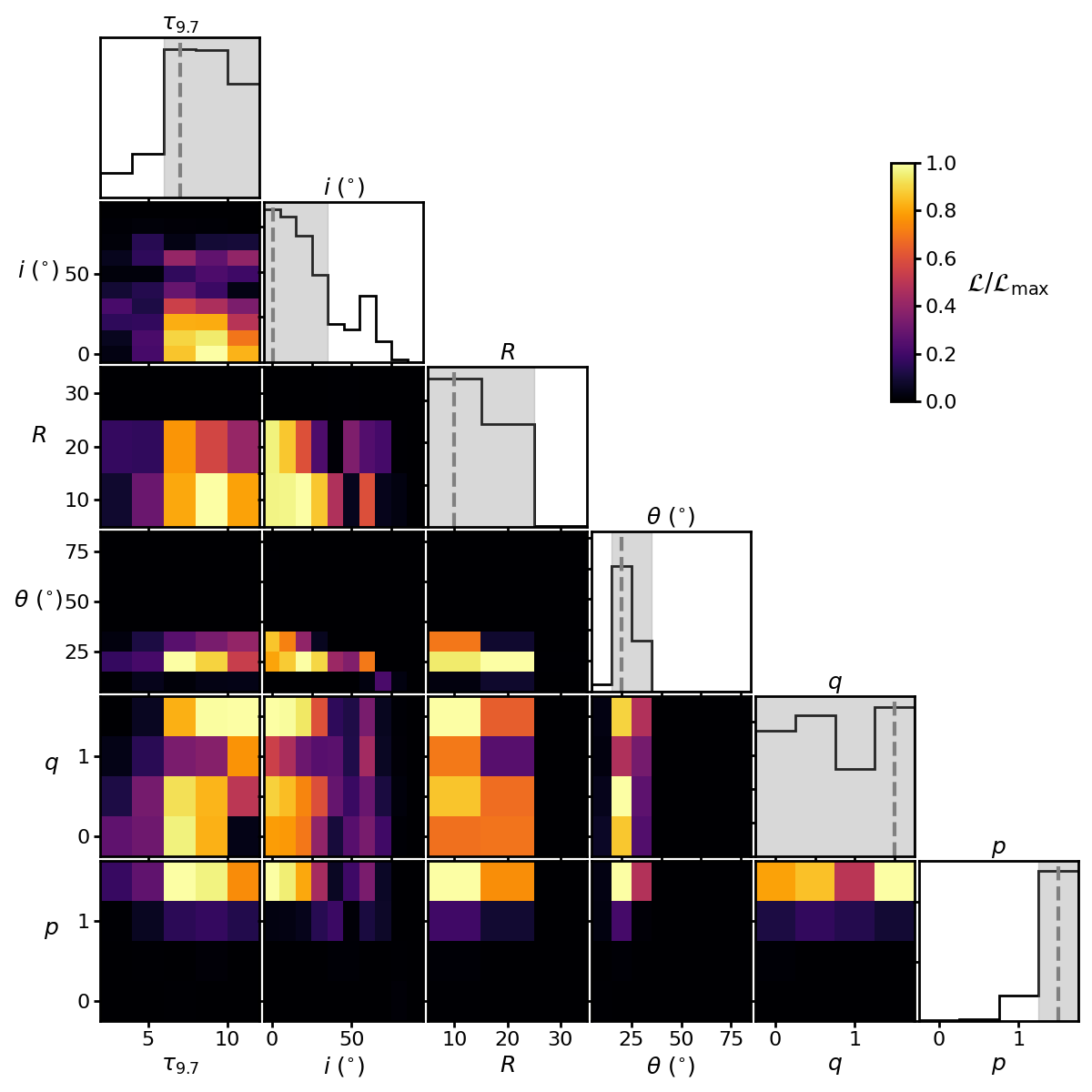}
\caption{Posterior likelihood of physical torus parameters for quasar J1342+0928, similar to Figure~\ref{fig:posterior}.}
\label{fig:app1}
\end{figure*}

\begin{figure*}
\centering
\includegraphics[width=0.79\textwidth]{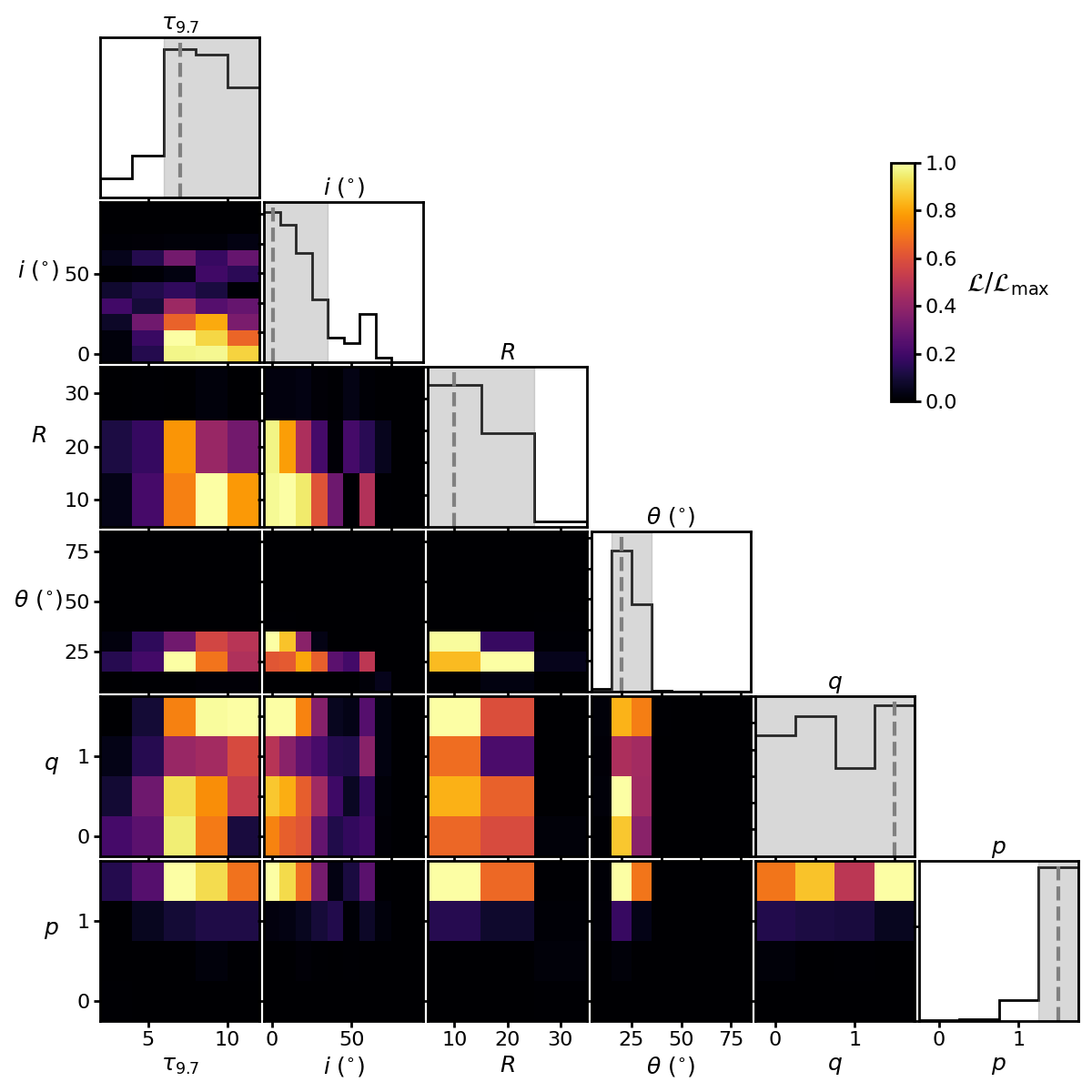}
\caption{Posterior likelihood of physical torus parameters for quasar J1007+2115, similar to Figure~\ref{fig:posterior}.}
\label{fig:app1}
\end{figure*}

\begin{figure*}
\centering
\includegraphics[width=0.79\textwidth]{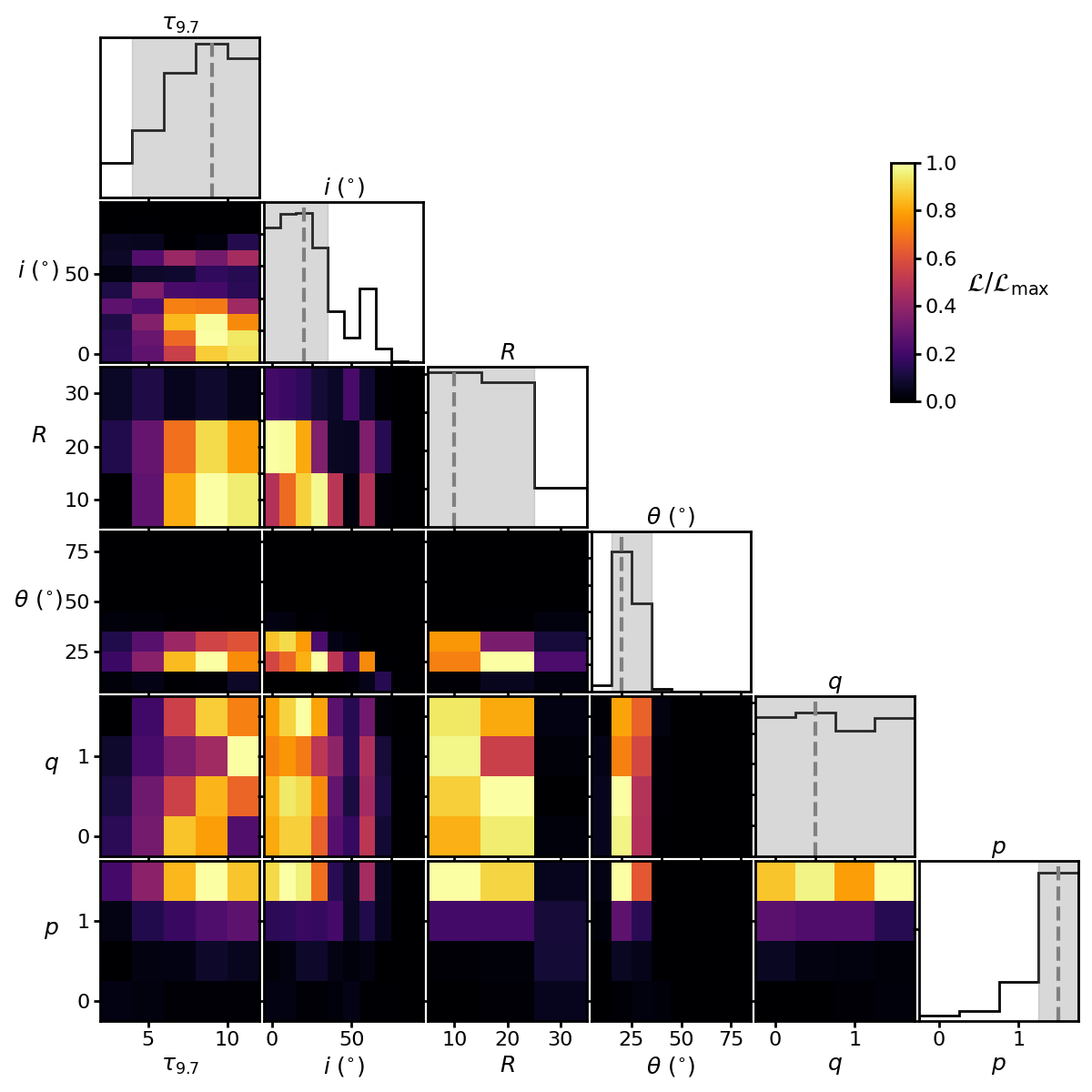}
\caption{Posterior likelihood of physical torus parameters for quasar J1120+0641, similar to Figure~\ref{fig:posterior}.}
\label{fig:app1}
\end{figure*}

\end{CJK}

\end{document}